\documentclass[12pt]{article}
\usepackage{graphicx,amsmath,amsfonts,  amssymb, amsthm, natbib,
  dsfont, tikz, color,algorithmic, algorithm, enumerate, enumitem,
  setspace, 
  blindtext}
\usetikzlibrary{positioning}
\usepackage[left=1in,top=1in,right=1in,bottom=1in,nohead,paperwidth=8.5in, paperheight=11in]{geometry} 
\newtheorem{proposition}{Proposition}[section]

\title{\bf Tree-based Node Aggregation in Sparse Graphical Models}
\author{Ines Wilms$^{a}$ and Jacob Bien$^b$
	\\ \textit{\small $^{a}$ Department of Quantitative Economics, Maastricht University, Maastricht, The Netherlands}
	\\ \textit{\small $^{b}$ Data Sciences and Operations, University of Southern California, Los Angeles, CA, USA}
}
\date{ }

\begin{document}
	\maketitle

\paragraph{Abstract.} 
High-dimensional graphical models are often estimated using
regularization that is aimed at reducing the number of edges in a network.
In this work, we show how even simpler networks can be produced by
aggregating the nodes of the graphical model. 
We develop a new convex regularized method, called the {\em tree-aggregated
  graphical lasso} or tag-lasso, that estimates graphical models
that are both edge-sparse and node-aggregated. 
The aggregation is performed in a data-driven fashion by leveraging side information in the form of a tree that encodes node similarity and facilitates the interpretation of the resulting aggregated nodes.
We provide an efficient implementation of the tag-lasso by using the
locally adaptive alternating direction method of multipliers and
illustrate our proposal's practical advantages in simulation and in applications in finance and biology.
\bigskip

\newpage

\section{Introduction} \label{intro}
Graphical models are greatly useful for understanding the relationships among large numbers of variables. 
Yet, estimating graphical models with many more parameters than
observations is challenging, which has led to an active area of
research on high-dimensional inverse covariance estimation. Numerous
methods attempt to curb the curse of dimensionality through regularized estimation procedures (e.g.,
\citealp{meinshausen2006high, yuan2007model, banerjee2008model,
  friedman2008sparse, rothman2008sparse, 
  peng2009partial, yuan2010high, 
  cai2011constrained, cai2016estimating}). 
Such methods aim for sparsity in the inverse covariance matrix, which corresponds
to graphical models with only a small number of edges.
A common method for estimating sparse graphical models is the graphical lasso (glasso) \citep{yuan2007model, banerjee2008model, rothman2008sparse, friedman2008sparse}, which adds an
$\ell_1$-penalty to the negative log-likelihood of a sample of multivariate
normal random variables.
While this and many other methods  
focus on the \textit{edges} for dimension reduction, far fewer contributions (e.g., \citealp{tan2015cluster, eisenach2020high, pircalabelu2020community})  focus on the {\em nodes} as a guiding principle for dimension reduction. 

Nonetheless, node dimension reduction is becoming increasingly 
relevant in many areas where 
data are being measured at finer levels of granularity. 
For instance,
in biology, modern high-throughput sequencing technologies provide
low-cost microbiome data at high resolution; 
in
neuroscience, brain activity in hundreds of regions of interest can be measured;
in finance, data
at the individual company level at short time scales are routinely analyzed;
and in marketing, joint purchasing data on every stock-keeping-unit (product) is recorded.
The fine-grained nature of this data brings new challenges. The sheer
number of fine-grained, often noisy, variables makes it difficult to
detect dependencies.  Moreover, there can be a mismatch between the
resolution of the measurement and the resolution at which natural
meaningful interpretations can be made.  The purpose of
an analysis may be to draw conclusions about entities at a coarser
level of resolution than happened to be measured.  Because of this mismatch,
practitioners are sometimes forced to devise ad hoc post-processing steps
involving, for example, coloring the nodes based on some classification of them into
groups  in
an attempt to make the structure of an estimated graphical model more
interpretable and the domain-specific takeaways more apparent (e.g., \citealp{millington2019quantifying}).

 \begin{figure}[t]
	\centering
	\includegraphics[width=\textwidth]{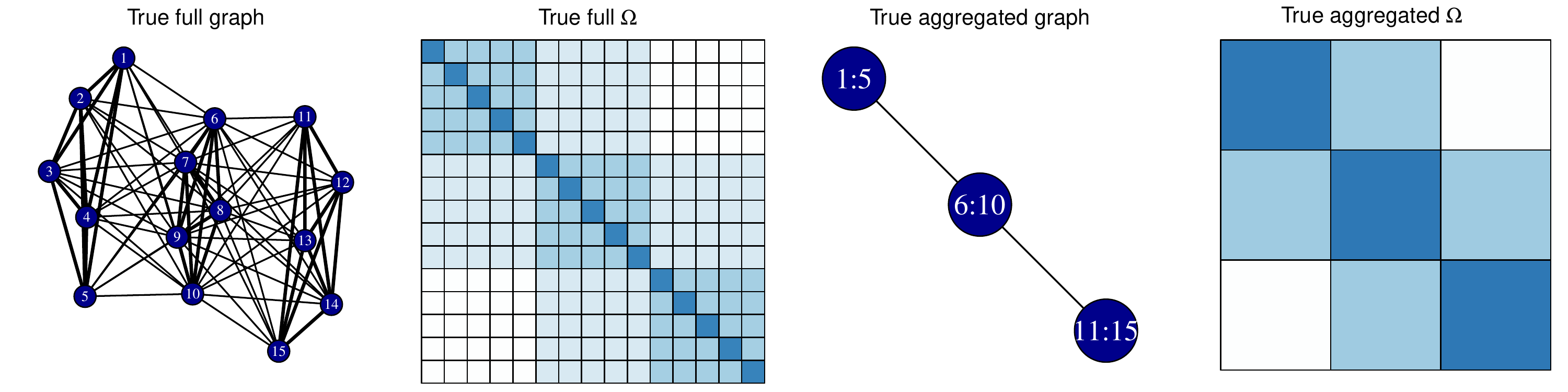}
	\includegraphics[width=\textwidth]{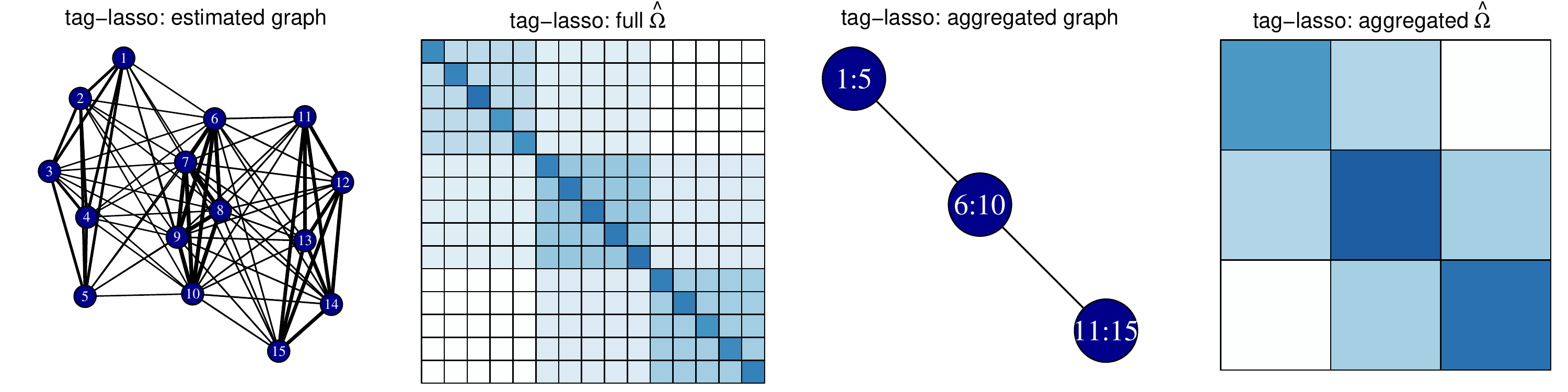}
	\includegraphics[width=\textwidth]{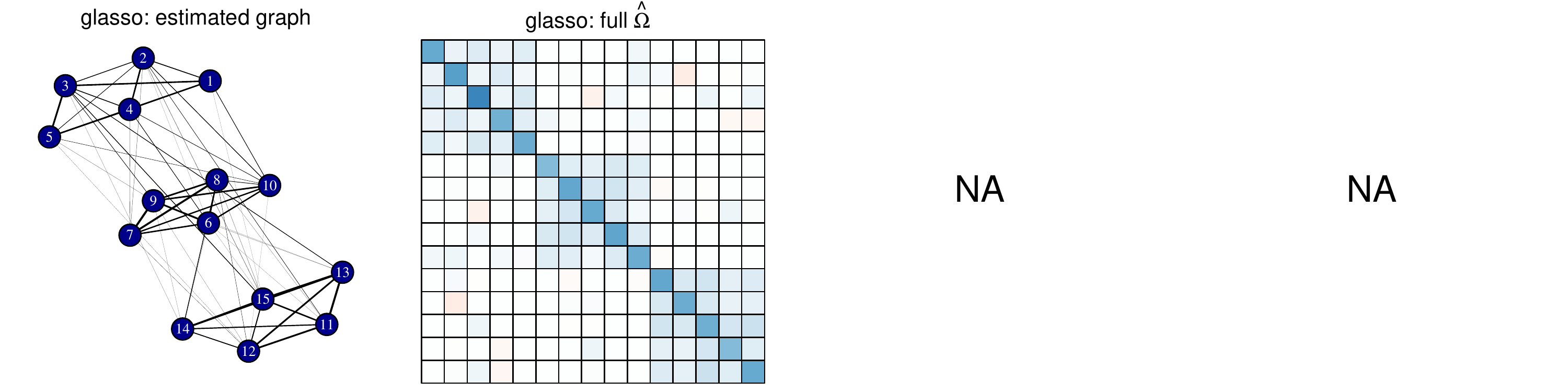}

\caption{ \label{intro_example} Top: True full graph and precision
  matrix $\boldsymbol\Omega$ with corresponding aggregated graph and precision
  matrix. Middle: Estimation output of the tag-lasso.  Bottom:
  Estimation output of the glasso. 
  }
\end{figure}	

Our solution to this problem is to incorporate the side information
about the relationship between nodes directly into the estimation
procedure.  In our framework, this side information is encoded as a
tree whose leaves correspond to the measured variables.  Such tree structures
are readily available in many domains (e.g., taxonomies in biology
and hierarchical classifications of jobs, companies, and products in business) and is well-suited to expressing
multi-resolution structure that is present in many problems.
We propose a new convex regularization procedure, called \textit{tag-lasso}, which stands for \textit{t}ree-\textit{a}ggregated-\textit{g}raphical-\textit{lasso}. 
This procedure combines node (or variable) aggregation with edge-sparsity.
The tree-based aggregation serves to both amplify the signal of
similar, low-level variables and render a graphical
model involving nodes at an appropriate level of scale to be relevant and interpretable. 
The edge-sparsity encourages the graphical model involving the
aggregated nodes has a sparse network structure.

Our procedure is based on a tree-based 
parameterization strategy that translates the node aggregation problem
into a sparse modeling problem, following an approach previously
introduced in the regression setting \citep{yan2018rare}. 
In Figure \ref{intro_example} (to be discussed more thoroughly in
Section \ref{simstudy}), we see that tag-lasso is able to recover the
aggregated, sparse graph structure.  By doing so, it yields a more
accurate estimate of the true graph, and its output is easier to interpret than the full, noisy graph obtained by the glasso.

The rest of the paper is organized as follows. 
Section \ref{node_aggregation} introduces the tree-based parameterization structure for nodewise aggregation in graphical models. 
Section \ref{method_ta_glasso} introduces the tag-lasso estimator, formulated as a solution to a convex optimization problem, for which we derive an efficient algorithm.
Section \ref{simstudy} presents the results of a simulation study. 
Section \ref{applications} illustrates the practical advantages of the
tag-lasso on financial and microbiome data sets.
Section \ref{conclusion} concludes.

\section{Node Aggregation in Penalized Graphical
  Models} \label{node_aggregation}
Let $\bf S$ be the empirical covariance matrix based on $n$
multivariate normal observations of dimension $p$, with mean vector
$\boldsymbol\mu$ and covariance matrix $\boldsymbol \Sigma$.  The target of
estimation is the precision matrix $\boldsymbol \Omega =
\boldsymbol \Sigma^{-1}$, whose sparsity pattern provides the graph structure of the
Gaussian graphical model, since $\Omega_{jk}=0$ is
equivalent to variables $j$ and $k$ being conditionally independent given all other variables.
To estimate the precision matrix, it is common to use a convex penalization method of the form
\begin{equation}
\widehat{\boldsymbol \Omega} = \underset{\boldsymbol \Omega}{\operatorname{argmin}} \{- \text{logdet}(\boldsymbol\Omega) + \text{tr}({\bf S}\boldsymbol{\Omega}) + \lambda \mathcal{P}(\boldsymbol\Omega) \ \ \text{s.t.} \  \boldsymbol{\Omega}=\boldsymbol{\Omega}^\top, \boldsymbol \Omega \succ 0\}, \label{PGM}
\end{equation}
where $\text{tr}(\cdot)$ denotes the trace, 
$\mathcal{P}(\cdot)$ is a convex penalty function, and $\lambda>$ is a tuning parameter controlling the degree of penalization.
Choosing the $\ell_1$-norm
\begin{equation}
\mathcal{P}(\boldsymbol\Omega) = \|\boldsymbol \Omega^{-\text{diag}}\|_1, \label{glasso_l1}
\end{equation}
where $\boldsymbol \Omega^{-\text{diag}}$ contains the unique
off-diagonal elements, yields the {\em graphical lasso} (glasso) \citep{friedman2008sparse,yuan2007model, banerjee2008model, rothman2008sparse}. 
It encourages $\widehat{\boldsymbol\Omega}$ to be sparse,
corresponding to a graphical model with few edges.

However, when $\boldsymbol\Omega$ is not sparse, demanding sparsity in
$\widehat{\boldsymbol\Omega}$ may not be helpful, as we will show in Section \ref{Sec2:node aggregation}. Such settings can arise when data are measured and analyzed at ever higher resolutions (a growing trend in many areas, see e.g. \citealt{callahan2017exact}).  
A tree is a natural way to represent the different scales of data resolution, and we introduce 
a new choice for $\mathcal{P}$ that uses this tree to guide node aggregation, thereby allowing for a data adaptive choice of data scale for capturing dependencies. 
Such tree-based structures are available in many domains. For instance, 
companies can be aggregated according to hierarchical industry classification codes; 
products can be aggregated from brands towards product categories; 
brain voxels can be aggregated according to brain regions;
microbiome data can be aggregated according to taxonomy.
The resulting penalty function then encourages a more
general and yet still highly interpretable structure for $\widehat{\boldsymbol\Omega}$.
In the following subsection, we use a toy example to illustrate the
power of such an approach.

\subsection{Node Aggregation} \label{Sec2:node aggregation}
Consider a toy example with $p$ variables
\begin{eqnarray}
X_1 & = & \sum_{j=3}^p X_j + \varepsilon_1 \nonumber \\
X_2 & = & \sum_{j=3}^p X_j + \varepsilon_2 \nonumber \\
X_j & = &  \varepsilon_j, \ \text{for} \ 3 \leq j \leq p, \nonumber 
\end{eqnarray}
where $\varepsilon_1,\ldots,\varepsilon_p$ are independent standard
normal random variables. 
By construction, it is clear that there is a very simple relationship
between the variables: The first two variables both depend on the sum
of the other $p-2$ variables.  However, a standard graphical model on the $p$
variables does not naturally express this simplicity.  The first row of Table
\ref{toy_example} shows the covariance and precision matrices for the
full set of variables $X_1, \ldots, X_p$.  The graphical model on the
full set of variables is extremely dense $O(p^2)$ edges.  Imagine if
instead we could form a graphical model with only three variables:
$X_1, X_2, \widetilde{X}$, where the last variable
$\widetilde{X}=\sum_{j=3}^p X_j$ aggregates all but the first two
variables.  The bottom row of Table \ref{toy_example} results in a
graphical model that matches the simplicity of the situation.

The lack of sparsity in the $p$-node graphical model means that the
graphical lasso will not do well.
Nonetheless, a method that could perform node aggregation would be able to yield a highly-interpretable aggregated sparse graphical model since $X_1$ and $X_2$ are conditionally independent given the aggregated variable  $\widetilde{X}$.

\begin{table}
\caption{Toy example: Covariance and precision matrices with
  corresponding graphical model (drawn for $p=50$) for the full (top) and aggregated (bottom) set of nodes. \label{toy_example}}
\begin{center}
	\resizebox{0.95\textwidth}{!}{\begin{minipage}{\textwidth}
		\begin{tabular}{cccc} \hline 
\textbf{Nodes} 			& \textbf{Covariance Matrix}  & \textbf{Precision Matrix}  &   \textbf{Graphical } \\
\textbf{}&$\boldsymbol{\Sigma}$&$\boldsymbol{\Omega}$&\textbf{Model}\\ \hline 
$X_1,  \ldots, X_p$&
			${\footnotesize\begin{pmatrix}
			p-1 & p-2 & {\bf 1}_{p-2}^\top \\ 
			p-2 & p-1 & {\bf 1}_{p-2}^\top \\ 
			{\bf 1}_{p-2}^\top & {\bf 1}_{p-2}^\top & {\bf I}_{p-2} \\ 
			\end{pmatrix}}$ & 
			${\footnotesize \begin{pmatrix}
			1 & 0 & -{\bf 1}_{p-2}^\top \\ 
			0 & 1 & -{\bf 1}_{p-2}^\top \\ 
			-{\bf 1}_{p-2} & -{\bf 1}_{p-2} &  {\bf L}\\ 
			\end{pmatrix}} $
			&
	        \raisebox{-1.5cm}{\includegraphics[scale=0.2]{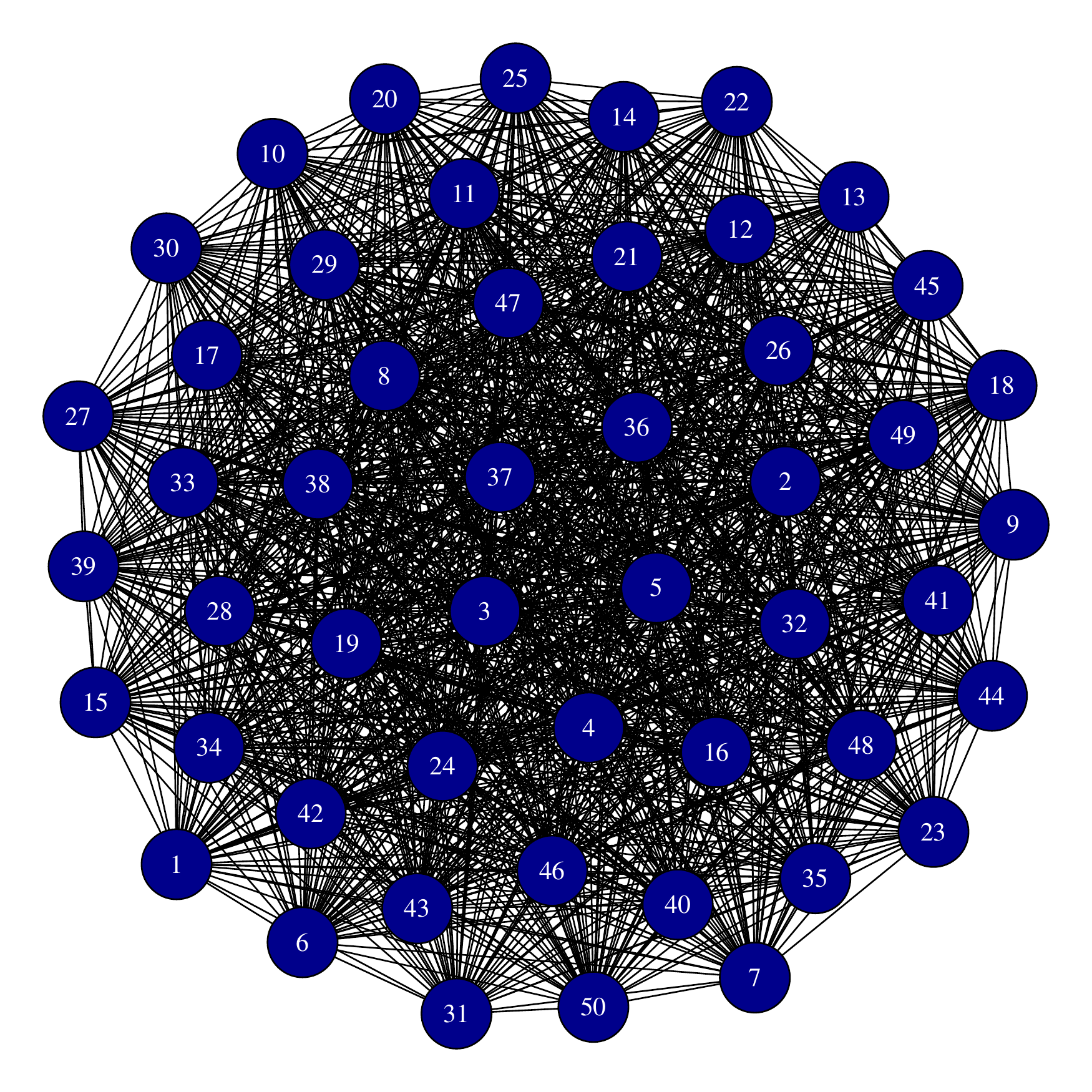}} \\	
			&& {\footnotesize with ${\bf L}={\bf I}_{p-2} + 2\cdot{\bf 1}_{p-2}\cdot{\bf 1}_{p-2}^\top$}&\\
&&&\\
$X_1, X_2, \widetilde{X}$ &			$ 
            {\footnotesize\begin{pmatrix}
			p-1 & p-2 & p-2 \\ 
			p-2& p-1 & p-2 \\ 
			p-2 & p-2 & p-2 \\
			\end{pmatrix}}$ & 
			${\footnotesize\begin{pmatrix}
			1 & 0 & -1 \\
		    0 & 1 & -1 \\
			-1 & -1 & 2+1/(p-2)
			\end{pmatrix}}$ & 
	        \raisebox{-1cm}{\includegraphics[scale=0.12]{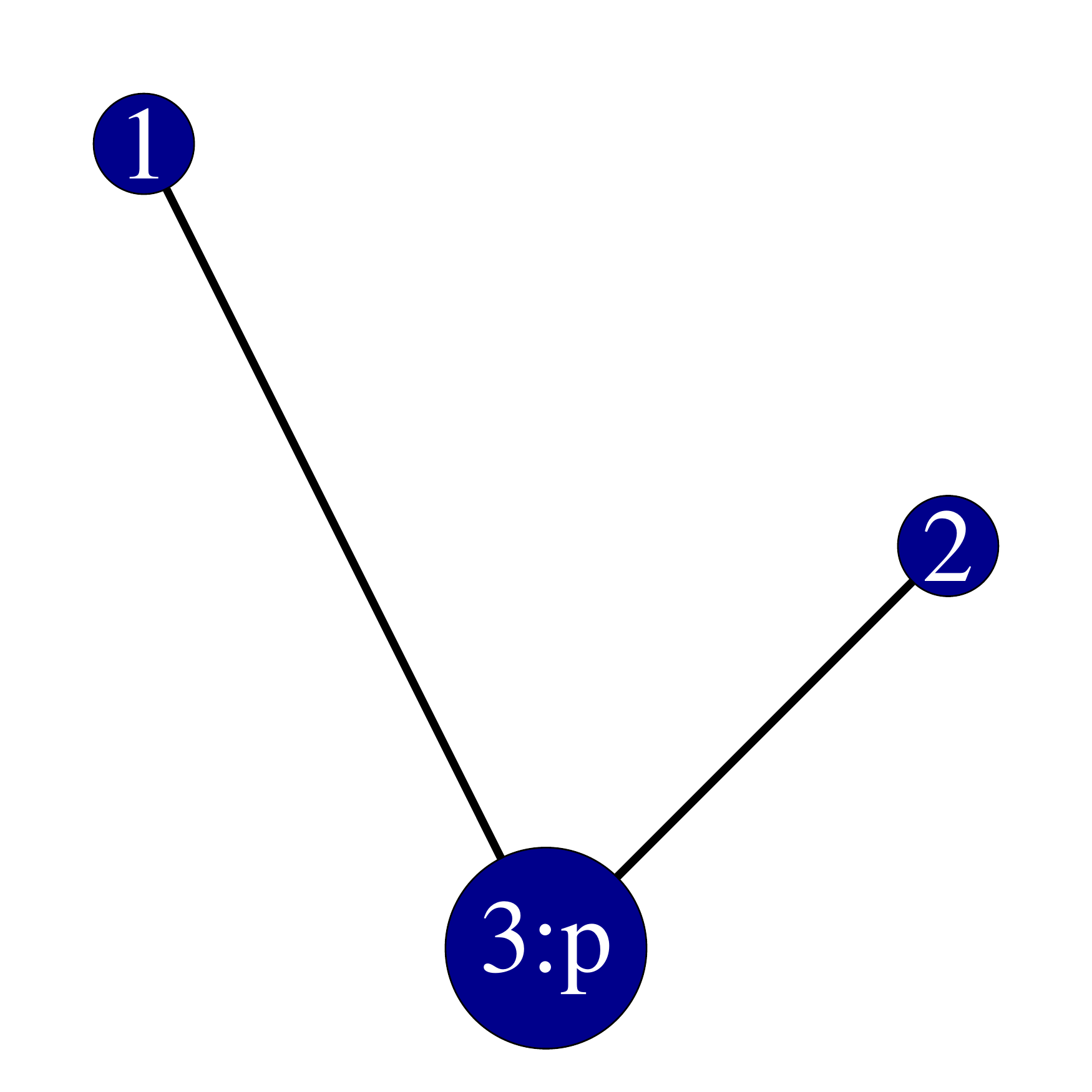}} \\
	        \hline
	        \multicolumn{4}{l}{Note: Let ${\bf 1}_d$ denote a $d$-dimensional column vector of ones, and ${\bf I}_d$ be the $d\times d$ identity matrix. }
			\end{tabular}
	\end{minipage} }
	\end{center}	
\end{table}

It is useful to map from the small aggregated graphical model to the
original $p$-node graphical model.  One does so by writing the precision matrix in ``$G$-block" format (\citealp{bunea2020model}, although they introduce this terminology in the context of the covariance matrix, not its inverse) for a given partition $G=\{G_1, ..., G_K\}$ of the nodes $\{1,\ldots, p\}$ and corresponding $p\times K$ membership matrix ${\bf M}$,  with entries $M_{jk}=1$ if $j \in G_k$, and $M_{jk}=0$ otherwise.
In particular, there exists a $K\times K$ symmetric  matrix ${\bf C}$ and  a $p\times p$  diagonal matrix  ${\bf D}$
such that the precision matrix can be written as 
$\boldsymbol\Omega= {\bf M} {\bf C} {\bf M}^\top +
{\bf D}$. 
The block-structure of $\boldsymbol \Omega$ is captured by the first part of the decomposition, the aggregated $K\times K$ precision matrix on the set of aggregated nodes can then be written as
$
\boldsymbol\Omega_{\text{agg}} = {\bf C} + {\bf D}_{\text{agg}},
$
where ${\bf D}_{\text{agg}} = ({\bf M}^\top {\bf D}^{-1}{\bf M})^{-1}$ is diagonal.
In the above example, $K=3$, $G_1 = \{1\}, \ G_2 = \{2\}, \ G_3 = \{3,
\ldots, p\}$ and 
${\bf M} {\bf C} {\bf M}^\top$ has only three distinct rows/columns
since the aggregated variables $j=3, \ldots, p$ share all their entries. In the presence of node aggregation and edge sparsity, the graphical model corresponding to the aggregated precision matrix is far more parsimonious than the graphical model on the full precision matrix (see Table \ref{toy_example}).

As motivated by this example, our main goal is to estimate the precision matrix in such a way that we can navigate from a $p$-dimensional problem to a $K$-dimensional problem whose corresponding graphical model provides a simple description of the conditional dependency structure among $K$ aggregates of the original variables. In the following proposition, we show that this can be accomplished by looking for a precision matrix that has a $G$-block structure. The proof of the proposition is included in Appendix \ref{app_proof_prop}

\begin{proposition} \label{omega_p_to_K}
Suppose ${\bf X} \sim N_p( {\bf 0}, \boldsymbol \Omega^{-1})$ with $\boldsymbol \Omega = {\bf M}{\bf C}{\bf M}^\top + {\bf D}$, 
where ${\bf M}\in\{0,1\}^{p\times K}$ is the membership matrix, $\bf D\succ 0$, and let ${\bf \widetilde{X}} = {\bf M}^\top {\bf X}\in\mathds{R}^K$ be the vector of aggregated variables. 
Then ${\bf \widetilde{X}}$ has precision matrix ${\bf C}+{\bf D}_{\text{agg}}$, where ${\bf D}_{\text{agg}}$ is a diagonal matrix, and therefore
$c_{ij} = 0$ is equivalent to  the aggregates $\widetilde{X}_i$ and $\widetilde{X}_j$ being conditionally independent given all other aggregated variables.
\end{proposition}

While Proposition \ref{omega_p_to_K} gives us the desired interpretation in the graphical model with $K$ aggregated nodes, 
\color{black}
in practice, the partition $G$, its size $K$, and corresponding
membership matrix ${\bf M}$ are, however, unknown. 
Rather than considering arbitrary partitions of the variables, we
constrain ourselves specifically to partitions guided by a known
tree.  In so doing, we allow ourselves to exploit side information and
help ensure that the aggregated nodes will be easily interpretable.
To this end, we introduce a tree-based parameterization strategy that allows us to embed the node dimension reduction  into a convex optimization framework.

\subsection{Tree-Based Parameterization} \label{Sec2:tree}
\begin{figure}
\centering
	\includegraphics[width=0.3\textwidth]{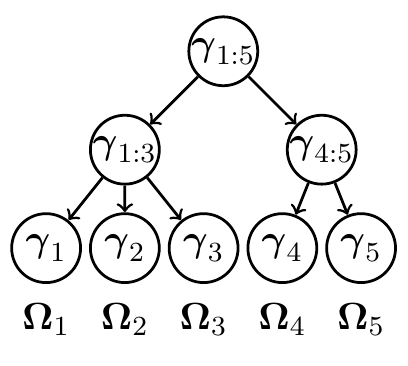}
	\caption{An example of a tree $\mathcal{T}$ encoding similarity among $p=5$ variables. \label{tree}}
\end{figure}

Our aggregation procedure assumes that we have, as side information, a tree that represents the closeness (or similarity) of variables. We introduce here a matrix-valued extension of the tree-based
parameterization developed in \citet{yan2018rare} for the regression setting.
We consider a tree $\mathcal{T}$ with $p$ leaves $\boldsymbol \Omega_1, \ldots, \boldsymbol \Omega_p$ where $\boldsymbol \Omega_j$ denotes column $1 \leq j \leq p$ of $\boldsymbol \Omega$. 
We restrict ourselves to partitions that can be expressed as a collection of branches of $\mathcal{T}$.
Newly aggregated nodes are then formed by summing variables within branches.
To this end, we assign a $p$-dimensional parameter vector
${\boldsymbol\gamma}_u$ to each node $u$ in the tree $\mathcal{T}$ (see Figure \ref{tree} for an example).
Writing  the set of nodes in the path from the root to the $j^{\text{th}}$ leaf (variable)  as $\text{ancestor}(j) \cup \{j\}$, 
we express each column/row in the precision matrix as
\begin{equation}
\boldsymbol{\Omega}_j = \sum_{u \in \text{ancestor}(j) \cup \{j\}} \boldsymbol \gamma_u + d_j {\bf e}_j, \label{Omaggregation}
\end{equation}
where we sum over all the $\boldsymbol \gamma_u$'s along this path, and
 ${\bf e}_j$ denotes the $p$-dimensional vector with all zeros except for its $j^{\text{th}}$ element that is equal to one.
In the remainder, we will make extensive use of the more compact notation
$ \boldsymbol \Omega = {\bf A}\boldsymbol{\Gamma} + {\bf D},$
where ${\bf A} \in \{0,1\}^{p \times |\mathcal{T}|}$ is a binary matrix with $A_{jk} = 1{\{u_k \in \text{ancestor}(j) \cup \{j\}\}} = 1{\{j \in \text{descendant}(u_k) \cup \{u_k\}\}}$,
$\boldsymbol\Gamma$ is a $|\mathcal{T}|\times p$  parameter matrix
collecting the $\boldsymbol\gamma_u$'s in its rows
and  ${\bf D}$ is a diagonal parameter matrix with elements $d_1, \ldots, d_p$. 

\begin{figure}
	\centering
	\includegraphics[width=\textwidth]{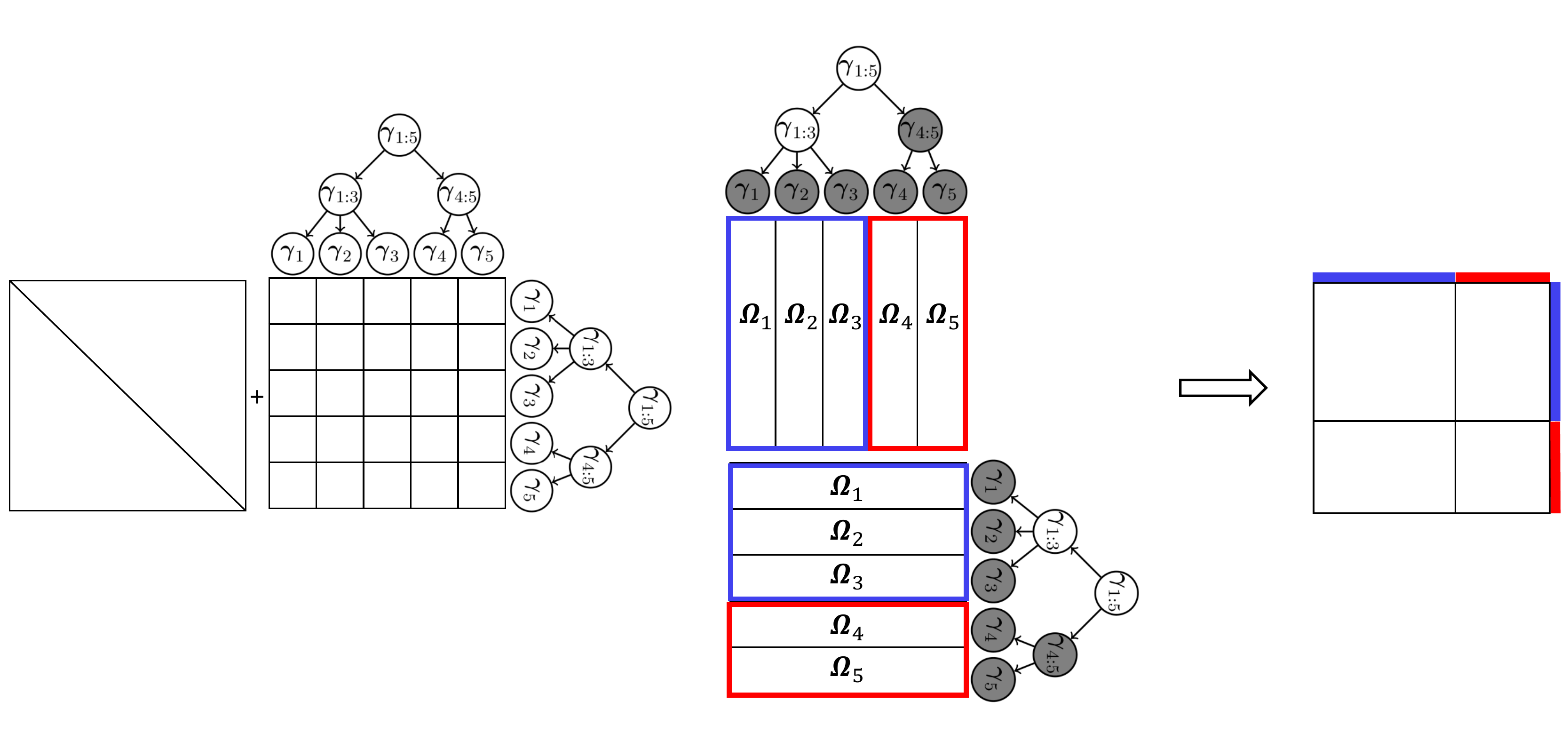}
	\vspace{-0.8cm}
	\caption{Left: An example of a $5 \times 5$-dimensional
          $\boldsymbol \Omega$ and a tree $\mathcal{T} $ that relates the corresponding $p=5$ variables. We have 
$\boldsymbol \Omega_{i} = {\boldsymbol \gamma}_i + \boldsymbol\gamma_{1:3} + \boldsymbol\gamma_{1:5}$ for $i=1,2,3$ and 
$\boldsymbol \Omega_{j} = {\boldsymbol \gamma}_j +
\boldsymbol\gamma_{4:5} + \boldsymbol\gamma_{1:5}$ for $j=4,5$, by equation \eqref{Omaggregation}, ignoring the diagonal elements.
Middle: By zeroing out the $\boldsymbol \gamma_i$'s in the gray nodes, we aggregate the rows/columns of $\boldsymbol \Omega$ into two groups indicated by the two colors:
$\boldsymbol \Omega_{1} = \boldsymbol \Omega_{2} = \boldsymbol \Omega_{3}  = \boldsymbol\gamma_{1:3} + \boldsymbol\gamma_{1:5}  $ (blue) and $\boldsymbol \Omega_{4} = \boldsymbol \Omega_{5} = \boldsymbol\gamma_{1:5}$ (red). Right: The precision matrix $\boldsymbol \Omega$ thus has a block-structure.} \label{tree_aggregation}
\end{figure}

By zeroing out $\boldsymbol \gamma_u$'s, certain nodes will be aggregated, as can be seen from the illustrative example in Figure \ref{tree_aggregation}. 
More precisely, let ${\mathcal{Z}} = \{u : {{\boldsymbol \gamma}}_u \neq {\bf 0} \}$ 
denote the set of non-zero rows in ${\boldsymbol \Gamma}$ and let ${\bf A}_{{\mathcal{Z}}}$ 
be the sub-matrix of ${\bf A}$ where only the columns corresponding to
the non-zeros rows in  ${{\boldsymbol \Gamma}}$ are kept. The number
of blocks $K$ in the aggregated network is then given by the  number
of unique  rows in ${\bf A}_{{\mathcal{Z}}}$. The membership matrix
$\bf M$ (Section \ref{Sec2:node aggregation}), and hence the set of aggregated nodes, can then be derived from the variables (rows) in the matrix ${\bf A}_{{\mathcal{Z}}}$ that share all their row-entries.

We are now ready to introduce the tag-lasso, which is  based on this parameterization.

\section{Tree Aggregated Graphical lasso} \label{method_ta_glasso}
To achieve dimension reduction via node aggregation and edge sparsity
simultaneously, we extend optimization problem \eqref{PGM} by
incorporating the parameterization introduced above. 
Our estimator, called the \textit{tag-lasso}, is defined as
\begin{multline}
(\widehat{\boldsymbol \Omega}, \widehat{\boldsymbol \Gamma}, \widehat{\boldsymbol D})= \underset{\boldsymbol \Omega, \boldsymbol \Gamma, \boldsymbol D}{\operatorname{argmin}} \{- \text{logdet}(\boldsymbol\Omega) + \text{tr}({\bf S}\boldsymbol{\Omega}) + \lambda_1 \|\boldsymbol{\Gamma}_{-r}\|_{2,1} +  \lambda_2 \|\boldsymbol \Omega^{-\text{diag}}\|_1\ \\ \text{s.t.} \  \boldsymbol{\Omega}=\boldsymbol{\Omega}^\top, \boldsymbol \Omega \succ {\bf 0},
\boldsymbol\gamma_r = \gamma {\bf 1}_p, \
\boldsymbol \Omega = {\bf A}\boldsymbol{\Gamma} + {\bf D},  \ {\bf D} \ \text{diag},  \ D_{jj} \geq 0 \ \text{for} \  j=1,\ldots,p \}, \label{ta-glasso}
\end{multline} 
with
$\|\boldsymbol{\Gamma}_{-r}\|_{2,1} = \sum_{u \in
  \mathcal{T}_{-r}}\|\boldsymbol \gamma_u\|_2$ and $\mathcal{T}_{-r}$
being the set of all nodes in $\mathcal T$ other than the root.
This norm induces row-wise sparsity on all non-root rows of ${\boldsymbol\Gamma}$.  
This row-wise sparsity, in turn, induces node aggregation as explained in Section \ref{Sec2:tree}. 
The root is excluded from this penalty term 
so that in the extreme of large $\lambda_1$ one gets complete
aggregation but not necessarily sparsity (in this extreme, all
off-diagonal elements of $\widehat{\boldsymbol\Omega}$ are equal to
the scalar $\gamma$ that appears in the equality constraint involving $\boldsymbol\gamma_r$).
While $\lambda_1$ controls the degree of node aggregation, $\lambda_2$
controls the degree of edge sparsity.
When $\lambda_1=0$, the optimization problem in \eqref{ta-glasso} reduces to the glasso.

Finally, note that optimization problem \eqref{ta-glasso} fits into the general formulation of penalized graphical models given in \eqref{PGM} since it can be equivalently expressed as 
$$
\widehat{\boldsymbol \Omega} = \underset{\boldsymbol \Omega}{\operatorname{argmin}} \{- \text{logdet}(\boldsymbol\Omega) + \text{tr}({\bf S}\boldsymbol{\Omega}) + \lambda_1 \mathcal{P}_\text{aggregate}(\boldsymbol\Omega) +  \lambda_2 \mathcal{P}_\text{sparse}(\boldsymbol\Omega) \  \text{s.t.} \  \boldsymbol{\Omega}=\boldsymbol{\Omega}^\top, \boldsymbol \Omega \succ {\bf 0}\}, $$
where 
$$
\mathcal{P}_\text{aggregate}(\boldsymbol\Omega) = \underset{\boldsymbol\Gamma, {\bf D}}{\operatorname{min}} \ \{\|\boldsymbol{\Gamma}_{-r}\|_{2,1} \ \text{s.t.} \
\boldsymbol\gamma_r = \gamma {\bf 1}_p, \
\boldsymbol \Omega = {\bf A}\boldsymbol{\Gamma} + {\bf D},  \ {\bf D} \ \text{diag},  \ D_{jj} \geq 0 \ \text{for} \  j=1,\ldots,p\}
$$
 and $\mathcal{P}_\text{sparse}(\boldsymbol\Omega)$ is the $\ell_1$-norm defined in \eqref{glasso_l1}.

\subsection{Locally Adaptive Alternating Direction Method of
  Multipliers}
\label{sec:la-admm}
We develop an {\em alternating direction method of multipliers} (ADMM) algorithm \citep{Boyd11}, specifically tailored to solving \eqref{ta-glasso}.
Our ADMM algorithm is based on solving this equivalent formulation of  \eqref{ta-glasso}: 
\begin{multline}
\underset{\underset{\boldsymbol\Gamma^{(1)}, \boldsymbol\Gamma^{(2)},\boldsymbol \Omega, \boldsymbol \Gamma, \boldsymbol D}{\boldsymbol\Omega^{(1)}, \boldsymbol\Omega^{(2)}, \boldsymbol\Omega^{(3)}}}{\operatorname{min}} \{- \text{logdet}(\boldsymbol\Omega^{(1)}) + \text{tr}({\bf S}\boldsymbol{\Omega}^{(1)}) + \lambda_1 \|\boldsymbol{\Gamma}^{(1)}_{-r}\|_{2,1} +  \lambda_2 \|\boldsymbol \Omega^{-\text{diag}(3)}\|_1 \\ \text{s.t.} \  \boldsymbol{\Omega}^{(1)}={\boldsymbol{\Omega}^{(1)}}^\top, 
\boldsymbol \Omega^{(1)} \succ {\bf 0}, 
\boldsymbol\gamma_r^{(1)} = \gamma^{(1)} {\bf 1}_p, \
\boldsymbol \Omega^{(2)} = {\bf A}\boldsymbol{\Gamma}^{(2)} + {\bf D}, 
 \ {\bf D} \ \text{diag},  \ D_{jj} \geq 0 \ \text{for} \  j=1,\ldots,p , \\
\boldsymbol{\Omega} = \boldsymbol{\Omega}^{(1)} = \boldsymbol{\Omega}^{(2)} =\boldsymbol{\Omega}^{(3)} \ \text{and} \ 
\boldsymbol{\Gamma} = \boldsymbol{\Gamma}^{(1)} = \boldsymbol{\Gamma}^{(2)} 
\}. \label{ADMMglobal}
\end{multline}
Additional copies of $\boldsymbol\Omega$ and $\boldsymbol\Gamma$ are introduced to efficiently decouple the optimization problem. 

Furthermore, we use an extension called {\em locally adaptive-ADMM} (LA-ADMM,
\citealp{xu2017admm}) with adaptive penalization to improve performance.
The full details of the algorithm are provided in Appendix
\ref{Appendix_ADMM}.  

\subsection{Selection of the Tuning Parameters}
To select the tuning parameters $\lambda_1$ and $\lambda_2$, we form
a $10\times 10$ grid of ($\lambda_1, \lambda_2$) values and find the
pair that minimizes a 5-fold
cross-validated likelihood-based score,
\begin{equation}
\frac1{5}\sum_{k=1}^5\left\{-
  \text{logdet}(\widehat{\boldsymbol\Omega}_{-\mathcal F_k}) + \text{tr}({\bf S}_{\mathcal F_k}\widehat{\boldsymbol{\Omega}}_{-\mathcal F_k})\right\}, \label{cv-error}
\end{equation}
where  $\widehat{\boldsymbol\Omega}_{-\mathcal F_k}$ is an estimate of
the precision matrix trained while withholding the samples in the $k^{\text{th}}$ fold and
${\bf S}_{\mathcal F_k}$ is the sample covariance matrix computed on the
$k^{\text{th}}$ fold.  In particular, we take
$\widehat{\boldsymbol\Omega}_{-\mathcal F_k}$ to be a re-fitted
version of our estimator (e.g., \citealp{belloni2013least}).
After fitting the tag-lasso, we obtain
$\widehat{\mathcal{Z}} = \{u : \widehat{{\boldsymbol \gamma}}_u \neq {\bf 0} \},$ the set of non-zero rows in $\widehat{{\boldsymbol \Gamma}}$, 
which suggests a particular node aggregation; and
$ \widehat{\mathcal{P}} = \{ (i,j) : \widehat{{\boldsymbol \Omega}}_{ij} \neq {\bf 0} \},$ the set of non-zero elements in 
$\widehat{{\boldsymbol \Omega}}$, 
which suggests a particular edge sparsity structure.
We then re-estimate $\boldsymbol{\Omega}$ by maximizing the likelihood subject to these aggregation and sparsity constraints:
\begin{equation}
\begin{aligned}
 &  \underset{\boldsymbol \Omega, \boldsymbol \Gamma_{\widehat{\mathcal{Z}}}, \boldsymbol D}{\operatorname{min}}
& & - \text{logdet}(\boldsymbol\Omega) + \text{tr}({\bf S}\boldsymbol{\Omega})   \\
& \text{subject to}
& & \boldsymbol{\Omega}=\boldsymbol{\Omega}^\top, \boldsymbol \Omega \succ {\bf 0}, \\
& & & \boldsymbol\gamma_{\widehat{\mathcal{Z}}, r} = \gamma {\bf 1}_p,\\
&&& \boldsymbol \Omega = {\bf A}_{\widehat{\mathcal{Z}}}\boldsymbol{\Gamma}_{\widehat{\mathcal{Z}}} + {\bf D}, {\bf D} \ \text{diag.},  \ D_{jj} \geq 0 \ \text{for} \  j=1,\ldots,p  \\
&&& \boldsymbol{\Omega}_{ij} = 0, \ \text{for} \ (i,j) \notin \widehat{\mathcal{P}}. 
\label{refit}
\end{aligned}
\end{equation}
We solve this with an LA-ADMM algorithm similar to what is described
in Section \ref{sec:la-admm} and Appendix  \ref{Appendix_ADMM}. 

\subsection{Connections to Related Work} \label{related_work}
Combined forms of dimension reduction in graphical models can be found in, amongst others, \cite{chandrasekaran2010latent, tan2015cluster, eisenach2020high, brownlees2020community, pircalabelu2020community}.

\cite{chandrasekaran2010latent} consider a blend of principal component analysis with graphical modeling by combining sparsity with a low-rank structure.
\cite{tan2015cluster} and \cite{eisenach2020high} both propose
two-step procedures that first cluster variables in an initial dimension reduction step and subsequently estimate a cluster-based graphical model. 
\cite{brownlees2020community} introduce partial correlation network models with community structures but rely on the 
sample covariance matrix of the observations to perform spectral clustering.
Our procedure differs from these works by introducing a single convex
optimization problem that simultaneously induces aggregation
and edge sparsity for the precision matrix. 

Our work is most closely related to \cite{pircalabelu2020community}
who estimate a penalized graphical model and simultaneously classify
nodes into communities. 
However, \cite{pircalabelu2020community} do
not use tree-based node-aggregation.
Our approach, in contrast, considers the tree $\mathcal{T}$ as an important part of the problem to help determine the extent of node
aggregation, and as a consequence the number of aggregated nodes
(i.e. clusters, communities or blocks) $K$, in a data-driven way
through guidance of the tree-based structure on the
nodes.

\section{Simulations} \label{simstudy}
We investigate the advantages of jointly exploiting node aggregation and edge sparsity in graphical models.
To this end, we compare the performance of the tag-lasso to two benchmarks:
\begin{enumerate}

    \item[(i)] {\em oracle:} The aggregated, sparse graphical model in
      \eqref{refit} 
        is estimated subject to the true aggregation and sparsity constraints. The oracle is only available for simulated data and serves as a ``best case" benchmark.
    \item[(ii)] {\em glasso}: This does not perform any aggregation (corresponding to the tag-lasso with $\lambda_1=0$). A sparse graph on the full set of variables is estimated.
The glasso is computed using the same LA-ADMM algorithm as detailed in Appendix  \ref{Appendix_ADMM}. The tuning parameter is selected from a  10-dimensional grid as the value that minimizes the 5-fold cross-validation likelihood-based score in equation \eqref{cv-error} with $\widehat{\boldsymbol\Omega}_{-\mathcal F_k}$ taken to be the glasso estimate.
\end{enumerate}

All simulations were performed using the \texttt{simulator} package \citep{bien2016simulator} in \texttt{R} \citep{Rcoreteam}.
We evaluate the estimators in terms of three performance metrics: estimation accuracy, aggregation performance, and sparsity recovery.
We evaluate \textit{estimation accuracy} by averaging over many
simulation runs the
Kullback-Leibler (KL) distance
$$\text{KL} = - \text{logdet}(\boldsymbol{\Sigma}\widehat{\boldsymbol\Omega}) + \text{tr}(\boldsymbol{\Sigma}\widehat{\boldsymbol\Omega}) - p  ,$$
where $\boldsymbol{\Sigma}=\boldsymbol\Omega^{-1}$ is the true covariance matrix.
Note that the KL distance is zero if the estimated precision matrix equals the true precision matrix.

To evaluate \textit{aggregation performance}, we use two measures: the Rand index \citep{rand1971objective} and the adjusted Rand index \citep{hubert1985comparing}. Both indices measure the degree of similarity between the true partition on the set of nodes ${1,\ldots,p}$ and the estimated partition. 
The Rand index ranges from zero to one, where one means that both partitions are identical.
The adjusted Rand index performs a re-scaling to account for the fact that random chance will cause some variables to occupy the same group.

Finally, to evaluate \textit{sparsity recovery}, we use the false positive and false negative rates
\begin{equation*}
\begin{aligned}
\text{FPR} &= \frac{\#\{(i,j): \widehat{\Omega}_{ij} \neq 0 \: \text{and} \: \Omega_{ij} = 0\}}{\#\{(i,j): \Omega_{ij} = 0\}} \ \ \text{and} \ \
\text{FNR} &= \frac{\#\{(i,j): \widehat{\Omega}_{ij} = 0 \: \text{and} \: \Omega_j \neq 0\}}{\#\{(i,j): \Omega_{ij} \neq 0\}}.
\end{aligned}
\end{equation*}
The FPR reports the fraction of truly zero components of the precision matrix that are estimated as nonzero. 
The FNR gives the fraction of truly nonzero components of the precision matrix that are estimated as zero.

\begin{figure}[t]
	\centering
	\includegraphics[width=\textwidth]{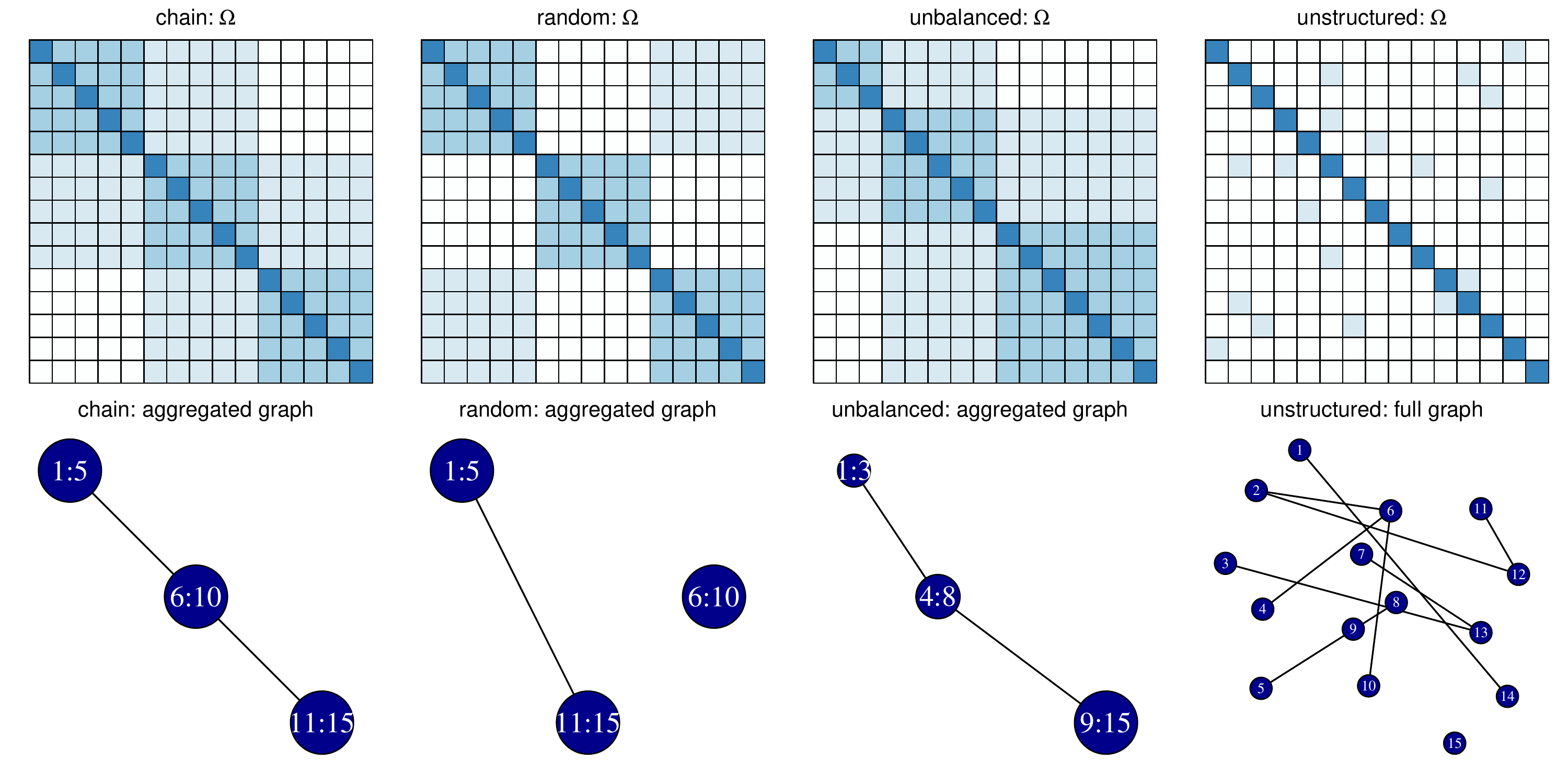}
	\vspace{-0.8cm}
\caption{ \label{sims_plots} Four aggregation designs: chain, random, unbalanced and unstructured graphs with corresponding precision matrix (top) and graph on the set of aggregated nodes (bottom).}
\end{figure}

\subsection{Simulation Designs} \label{sim:sec:design}
Data are drawn from a multivariate normal distribution with mean zero and covariance matrix $\boldsymbol\Sigma=\boldsymbol\Omega^{-1}$. 
We take $p=15$ variables and investigate the effect of increasing the number of variables in Section \ref{sim:sec:increasep}.
We consider four different simulation designs, shown in Figure \ref{sims_plots}, each having a different combination of aggregation and sparsity structures for the precision matrix $\boldsymbol\Omega$.

Aggregation is present in the first three structures. The precision matrix has a $G$-block structure with $K=3$ blocks. In Section \ref{sim:sec:increaseK}, we investigate the effect of varying the number of blocks.
In the \textit{chain} graph, adjacent aggregated groups are connected through an edge.
This structure corresponds to the motivating example of Section \ref{intro}.
In the \textit{random} graph, one non-zero edge in the aggregated network is chosen at random.
In the \textit{unbalanced} graph, the clusters are of unequal size. 
In the \textit{unstructured} graph, no aggregation is present. 

Across all designs, we take the diagonal elements of $\boldsymbol \Omega$ to be
$1$, the elements within a block of aggregated variables to be $0.5$,
and the non-zero elements across blocks to be $0.25$. 
We generate $100$ different data sets for every simulation design and use a sample size of $n=120$. The number of parameters ($p+p(p-1)/2 = 120$) equals the sample size.

\begin{figure}
	\centering
	\includegraphics[width=0.45\textwidth]{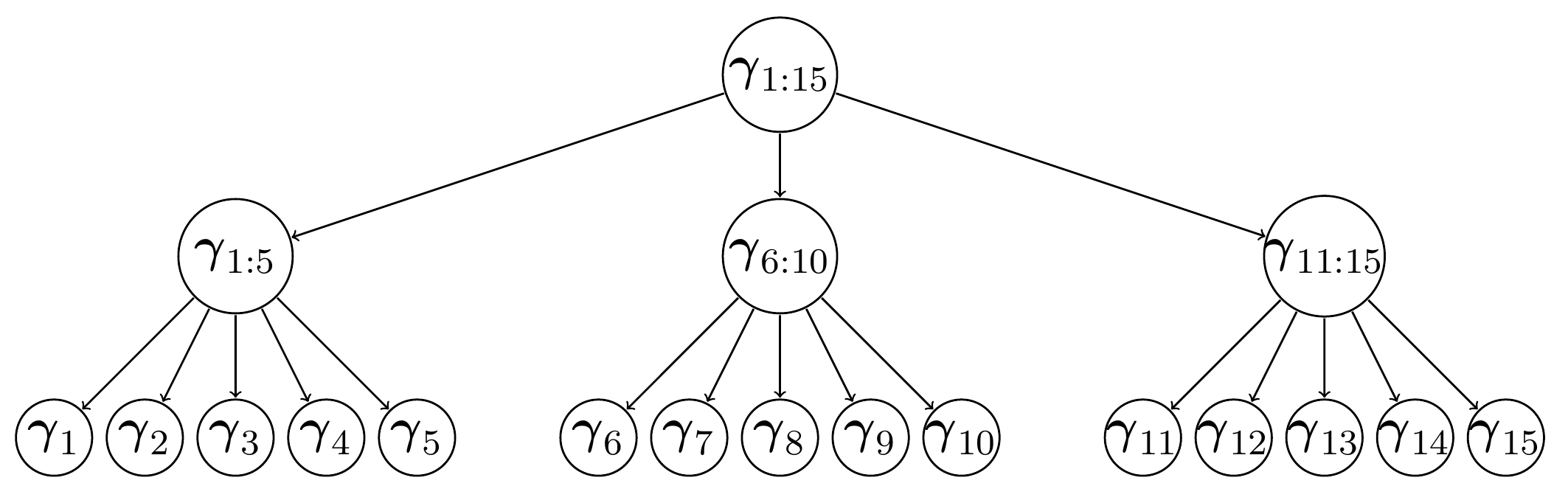}
	\hspace{0.8cm}
	\includegraphics[width=0.45\textwidth]{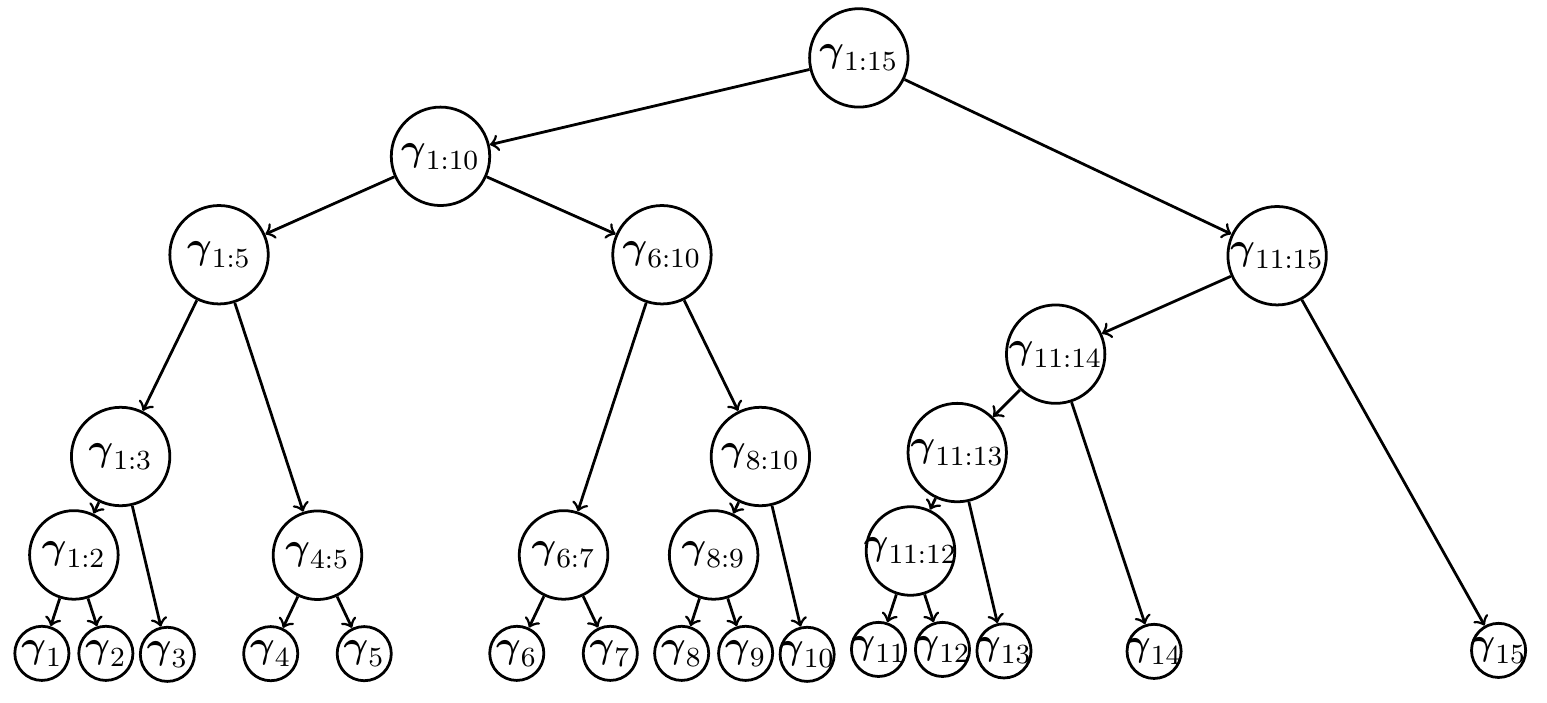}
\caption{ \label{sims_tree} A simple tree used for the ``tag-lasso
  ideal" (left) and a more realistic tree used for the ``tag-lasso realistic" (right).}
\end{figure}

The tag-lasso estimator relies on the existence of a tree to perform node dimension reduction. We consider two different tree structures throughout the simulation study.
First, we use an ``ideal" tree which contains the true aggregation
structure as the sole aggregation level between the leaves and the root of the tree.
As an example, the true aggregation structure for the chain graph
structure is shown in the left panel of Figure \ref{sims_tree}. 
We form ${\bf A}$ corresponding to this oracle tree to obtain the \textit{``tag-lasso ideal"} estimator.

We also consider a more realistic tree, shown in the right panel of Figure
\ref{sims_tree}, following a construction similar to that of
\citet{yan2018rare}.  The tree is formed by performing hierarchical clustering of $p$ latent points chosen to ensure that the tree contains the
true aggregation structure and that these true clusters occur across a variety of depths.  In particular,  we generate $K$ cluster means $\mu_1, \ldots, \mu_K$ with $\mu_i=1/i$.
We set the number of latent points associated with each of the $K$ means equal to the cluster sizes from Figure \ref{sims_plots}.
These latent points are then drawn independently from $N(\mu_i,[0.05\cdot\min_j(\mu_i - \mu_j)]^2)$.
Finally, we form ${\bf A}$ corresponding to this tree to obtain the \textit{``tag-lasso realistic"} estimator.

\subsection{Results} \label{sim:sec:results}
We subsequently discuss the results on estimation accuracy, aggregation performance, and sparsity recovery.

\begin{figure}
	\centering
	\includegraphics[width=\textwidth]{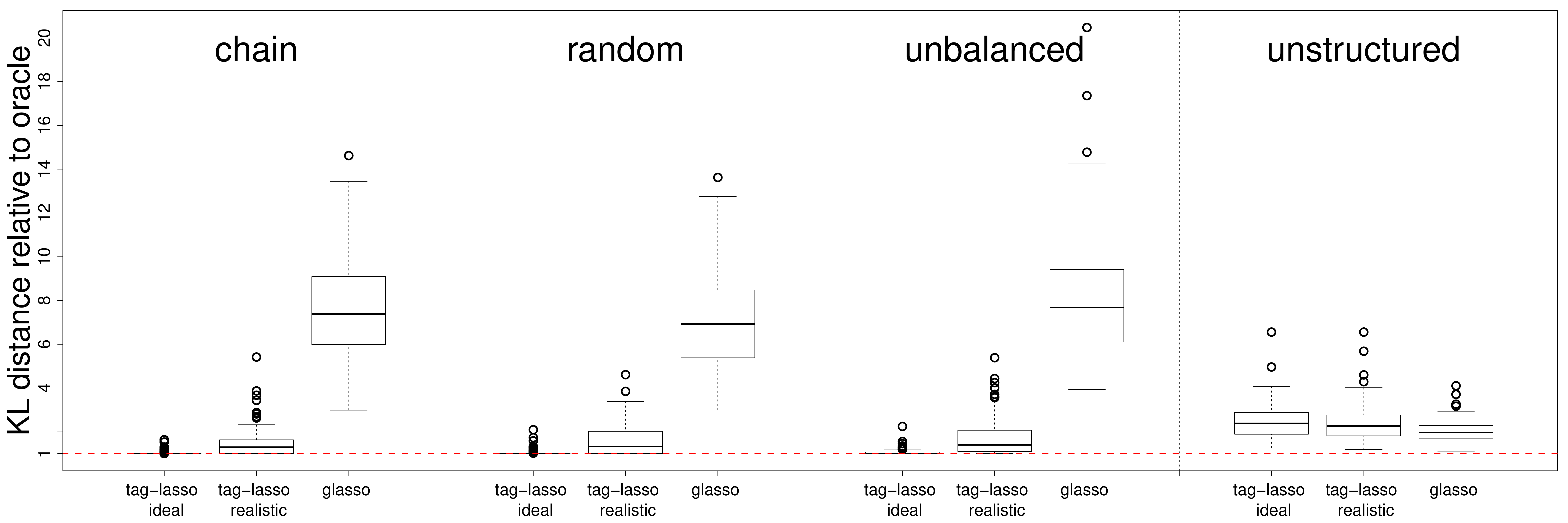}
\caption{ \label{sims_frob_p15} Estimation accuracy of the tree estimators relative to the oracle.
}
\end{figure}

\paragraph{Estimation Accuracy.} Boxplots of the  KL distances for the
three estimators (tag-lasso ideal, tag-lasso realistic and glasso) relative to
the oracle are  given in Figure \ref{sims_frob_p15}. 
The first three panels correspond to simulation designs with
aggregation structures. In these settings, the tag-lasso estimators considerably outperform the glasso, on average by a factor five. 
The tag-lasso ideal method performs nearly as well as the oracle. 
Comparing the tag-lasso realistic method to the tag-lasso ideal method suggests a minimal
price paid for using a more realistic tree. 

The ``unstructured'' panel of Figure \ref{sims_frob_p15} shows a case
in which
there is sparsity but no aggregation in the true data generating model. 
As expected, the glasso performs best in this case; however, we 
observe minimal cost to applying the tag-lasso approaches (which encompass the glasso as a special case when $\lambda_1 = 0$).

\paragraph{Aggregation Performance.} Table \ref{sims_agg_p15}
summarizes the aggregation performance of the three estimators in
terms of the Rand index (RI) and adjusted Rand index (ARI). No results
on the ARI in the unstructured simulation design are reported since 
it
cannot be computed for a partition consisting of
singletons. 
The tag-lasso estimators perform very well. If one can rely on an oracle tree, the tag-lasso perfectly recovers the aggregation structure, as reflected in the perfect (A)RI values of the tag-lasso ideal method.
Even when the tag-lasso uses a more complex tree structure, it recovers the correct aggregation structure in the vast majority of cases. 
The glasso returns a partition of singletons as it is unable to perform dimension reduction through aggregation, as can be seen from its zero values on the ARI.

\begin{table}
\caption{Aggregation performance of the three estimators, as measured
  by the Rand index (RI) and adjusted Rand index (ARI), for the four
  simulation designs.
  Standard errors are in
  parentheses. \label{sims_agg_p15}}
  \resizebox{0.85\textwidth}{!}{\begin{minipage}{\textwidth}
\begin{tabular}{lcccccccc} \hline 
Estimators & \multicolumn{2}{c}{chain} & \multicolumn{2}{c}{random} & \multicolumn{2}{c}{unbalanced} & \multicolumn{2}{c}{unstructured} \\ 
                    & RI & ARI & RI & ARI & RI & ARI & RI & ARI\\  \hline 
tag-lasso ideal    & 1.00 (.00) & 1.00 (.01) & 1.00 (.00) & 1.00 (.00) & 1.00 (.00)& 0.99 (.01) & 0.84 (.02)  & NA\\
tag-lasso realistic           & 0.95 (.01) & 0.88 (.01) & 0.97 (.01) & 0.93 (.01) & 0.94 (.01) & 0.85 (.02)  & 0.81 (.02) & NA \\
glasso              & 0.71 (.00) & 0.00 (.00) & 0.71 (.00) & 0.00 (.00) & 0.67 (.00) & 0.00 (.00)  & 1.00 (.00) & NA\\ \hline 
\end{tabular}
\end{minipage}}
\end{table}

\paragraph{Sparsity Recovery.} Table \ref{sims_sparsity_p15} summarizes the results on sparsity recovery  (FPR and FNR).
The tag-lasso estimators enjoy favorable FPR and FNR, mostly excluding the irrelevant conditional dependencies (as reflected by their low FPR) and including the relevant
conditional dependencies (as reflected by their low FNR). 
In the simulation designs with aggregation, the glasso pays a big
price for  not being able to reduce dimensionality through
aggregation, leading it to include too many irrelevant conditional
dependencies, as reflected through its large FPRs. 
In the unstructured design, the rates of all estimators are, overall, low. 

\begin{table}
\caption{Sparsity recovery of the three estimators, as measured by the
  false positive rate (FPR) and false negative rate (FNR), for the
  four simulation designs. Standard errors are in parentheses. \label{sims_sparsity_p15}}
\resizebox{0.82\textwidth}{!}{\begin{minipage}{\textwidth}
\begin{tabular}{lcccccccc} \hline 
Estimators & \multicolumn{2}{c}{chain} & \multicolumn{2}{c}{random} & \multicolumn{2}{c}{unbalanced} & \multicolumn{2}{c}{unstructured} \\ 
                    & FPR & FNR & FPR & FNR & FPR & FNR & FPR & FNR \\  \hline 
tag-lasso ideal    &  0.22 (.04) &  0.00 (.00) &  0.19 (.04) &  0.00 (.01) &  0.46 (.05) &  0.00 (.00) &  0.06 (.01) &  0.15 (.01) \\
tag-lasso realistic          &  0.30 (.04) &  0.02 (.01) &  0.13 (.02) &  0.09 (.01) &  0.44 (.04) &  0.05 (.01) &  0.05 (.01) &  0.14 (.01) \\
glasso              &  0.80 (.02) &  0.08 (.01) &  0.73 (.01) &  0.09 (.01) & 0.82 (.02) &  0.07 (.01) &  0.16 (.01) &  0.04 (.01) \\ \hline 
\end{tabular}
\end{minipage}}
\end{table}

\subsection{Increasing the Number of Nodes}  \label{sim:sec:increasep}
We investigate the sensitivity of our results to an increasing number of variables $p$. We focus on the chain simulation design from Section  \ref{sim:sec:design} and subsequently double $p$ from 15 to 30, 60 and 120 while keeping the number of blocks $K$ fixed at three.
The sample size $n$ is set proportional to the complexity of the model, as measured by $Kp + p$.
  Hence, the sample sizes corresponding to the increasing values of $p$ are respectively, $n=120,240,480,960$, thereby keeping the ratio of the sample size to the complexity fixed at two.
In each setting, the number of parameters to be estimated is large, equal to 120, 465, 1830, 7260, respectively; thus increasing relative to the sample size. 

\begin{figure}
	\centering
	\includegraphics[width=0.45\textwidth]{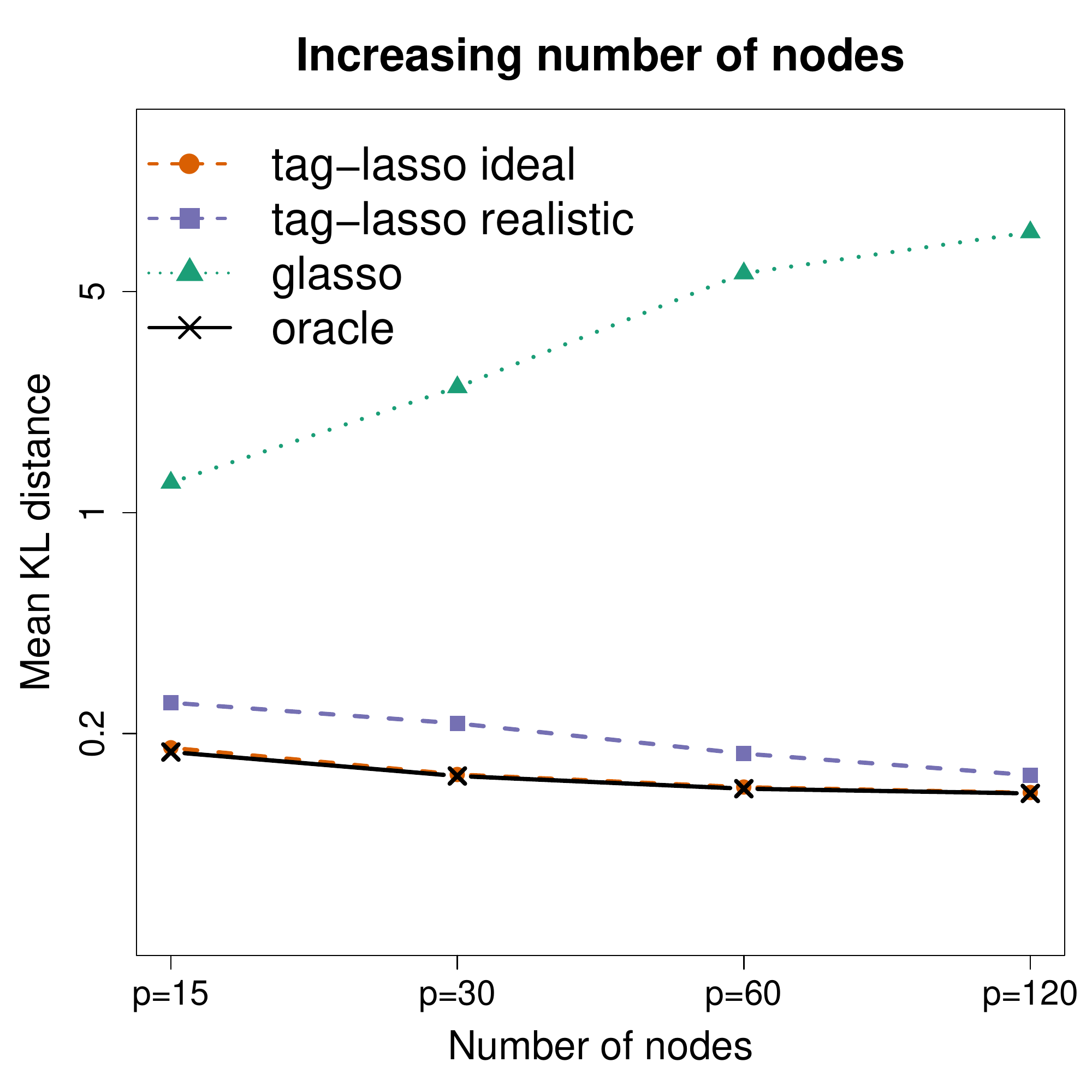}
	\hspace{1cm}
	\includegraphics[width=0.45\textwidth]{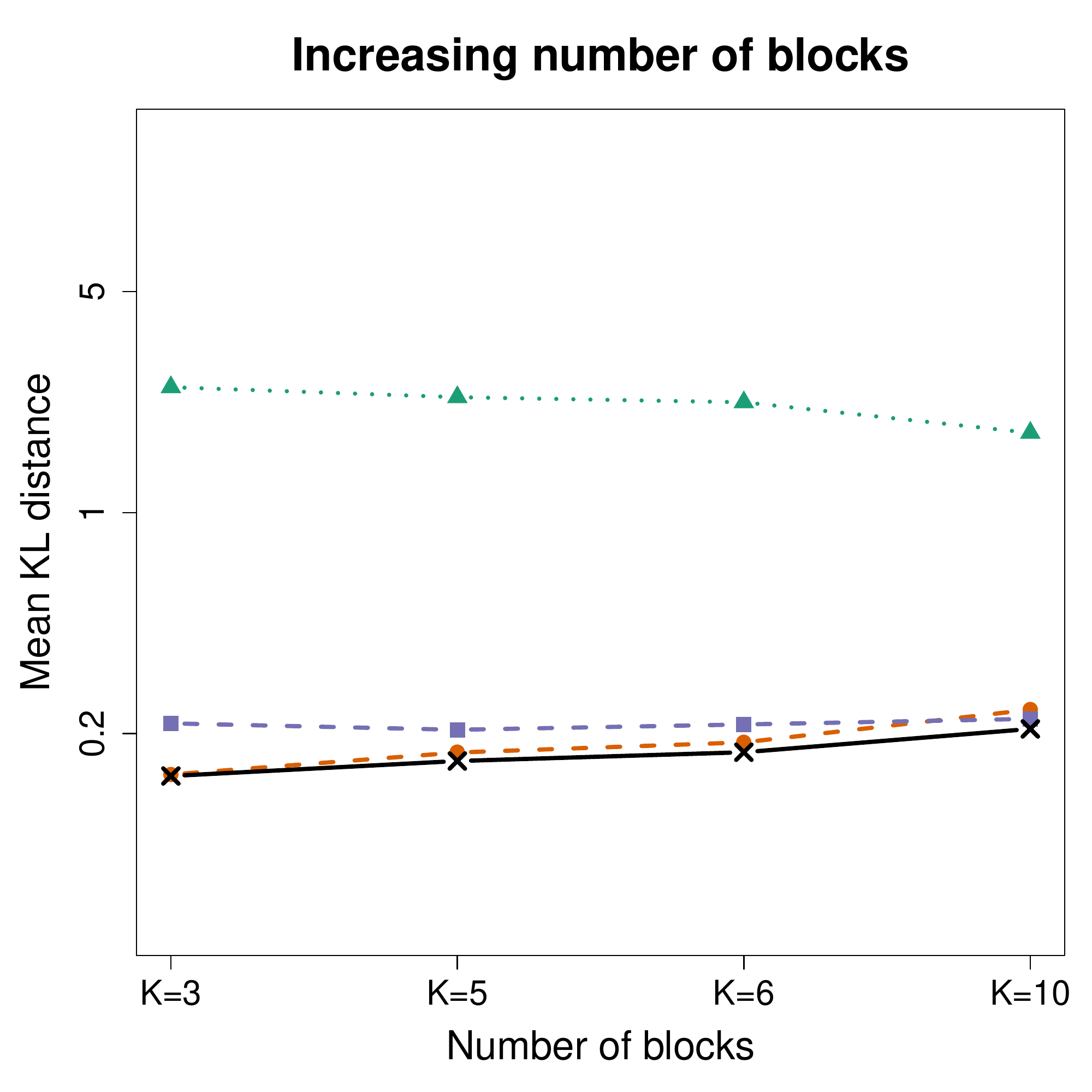}
\caption{ \label{sims_increasek} Estimation accuracy of the four estimators (on a log-scale) for increasing
number of variables $p$  (and fixed $K=3$, left panel)
the number of blocks $K$ (and fixed $p=30$, right panel). }
\end{figure}

The left panel of Figure \ref{sims_increasek} shows the mean KL distance (on a log-scale) 
of the four estimators as a function of $p$. 
As the number of nodes increases, the estimation accuracy of the
tag-lasso estimators and the oracle increases slightly. For fixed $K$
and increasing $p$, the aggregated nodes---which can be thought of as
the average of $p/K$ random variables---may be stabler, thereby
explaining why the problem at hand does not get harder when increasing
$p$ for the methods with node aggregation. By contrast, the
glasso---which is unable to exploit the aggregation structure---performs worse as $p$ increases. 
For $p=120$, for instance, the tag-lasso estimators outperform the
glasso by a factor 50.

Results on aggregation performance and sparsity recovery are presented in Figure \ref{sims_increasep_agg_and_sparsity} of  Appendix \ref{Appendix_sims}.
The tag-lasso ideal method perfectly recovers the aggregation structure for
all values of $p$. 
The realistic tag-lasso's aggregation performance is close to perfect and remains relatively stable as $p$ increases.
The glasso is unable to detect the aggregation structure, as expected and reflected through its zero ARIs.
The tag-lasso estimators also maintain a better balance between the FPR and FNR than the glasso. 
While their FPRs increase as $p$ increases, their FNRs remain close to perfect, hence all relevant conditional dependencies are recovered. The glasso, in contrast, fails to recover the majority of relevant conditional dependencies when $p=60,120$, thereby explaining its considerable drop in estimation accuracy.

\subsection{Increasing the Number of Blocks} \label{sim:sec:increaseK}
Finally, we investigate the effect of increasing the number of blocks $K$.
We take the chain simulation design from Section  \ref{sim:sec:design} and increase the number of blocks from $K=3$ to $K=5,6,10$, while keeping the number of variables fixed at $p=30$.
The right panel of Figure \ref{sims_increasek} shows the mean KL distance (on a log-scale)  of the four estimators as a function of $K$.
As one would expect, the difference between the aggregation methods
and the glasso decreases as $K$ increases.  However, for all $K$
considered, the glasso does far less well than the aggregation based methods.

Similar conclusions hold in terms of aggregation and sparsity recovery performance. Detailed results are presented in Figure \ref{sims_increasek_agg_and_sparsity} of  Appendix \ref{Appendix_sims}. 
The tag-lasso ideal method performs as well as the oracle in terms of
capturing the aggregation structure; the tag-lasso realistic method performs close to perfect and its aggregation performance improves with increasing $K$. 
In terms of sparsity recovery, 
the tag-lasso estimators hardly miss relevant conditional dependencies and only include a small number  of irrelevant conditional dependencies. 
The glasso's sparsity recovery performance is overall worse but does improve with increasing $K$.

\section{Applications} \label{applications}

\subsection{Financial Application} \label{finance}
\begin{figure}
	\centering
	\includegraphics[width=1\textwidth]{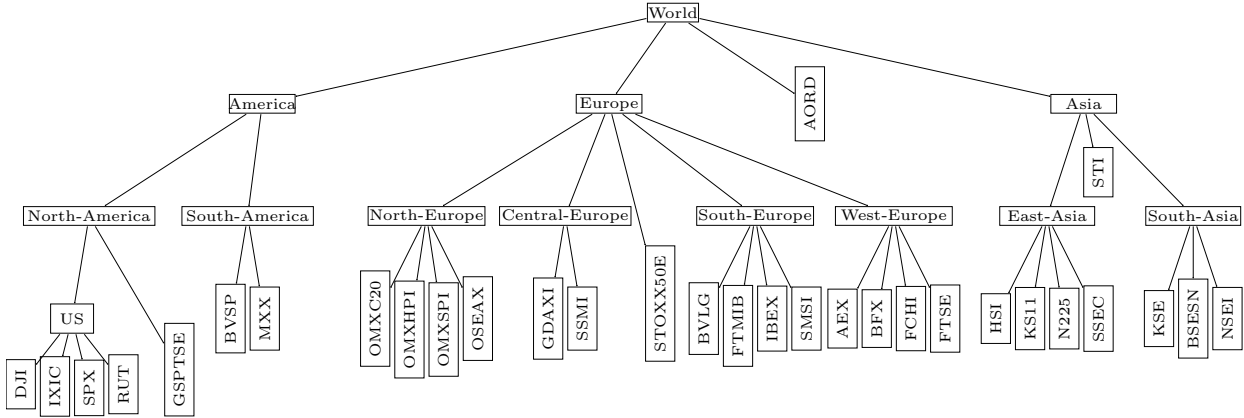}
\caption{\label{stock_market_tree}Geography-based tree for the stock
  market data, which aggregates the $p=31$ stock market indices
  (leaves) over several sub-continents towards a single root.  Leaves,
  which represent individual stock markets, are displayed
  horizontally.}
\end{figure}

We demonstrate our method on a financial data set containing daily
realized variances of $p=31$ stock market indices from across the world in 2019 ($n=254$). 
Daily  realized variances based on five minute returns 
are taken from the Oxford-Man Institute of 
Quantitative Finance (publicly available at http://realized.oxford-man.ox.ac.uk/data/download). Following standard practice,  all realized variances are  log-transformed. 
An overview of the stock market indices is provided in Appendix \ref{Appendix_Finance}. 
We encode similarity between the 31 stock market indices according to geographical region, and use the tree shown in Figure \ref{stock_market_tree} to apply the tag-lasso estimator.

Since the different observations of the consecutive days are
(time)-dependent, we first fit the popular and simple {\em heterogeneous autoregressive} (HAR) model  of \citep{Corsi09} to each of the individual log-transformed realized variance series. Graphical displays of the residual series of these 31 HAR models suggest that almost all autocorrelation in the series is captured. We then apply the tag-lasso to the residual series to learn the conditional dependency structure among stock market indices.

\begin{figure}[t]
\centering
\includegraphics[width=0.42\textwidth]{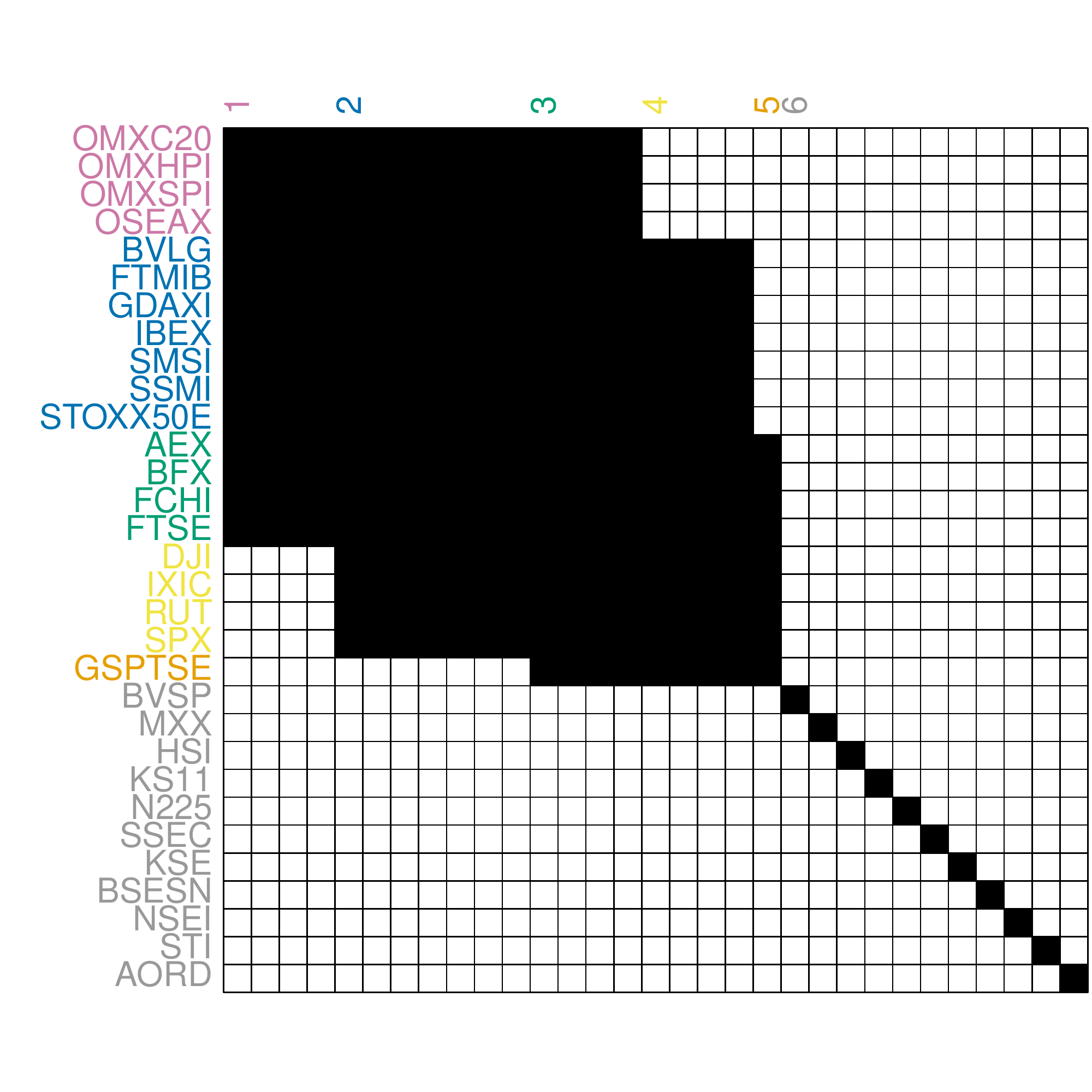}
\hspace{1cm}
\includegraphics[width=0.4\textwidth]{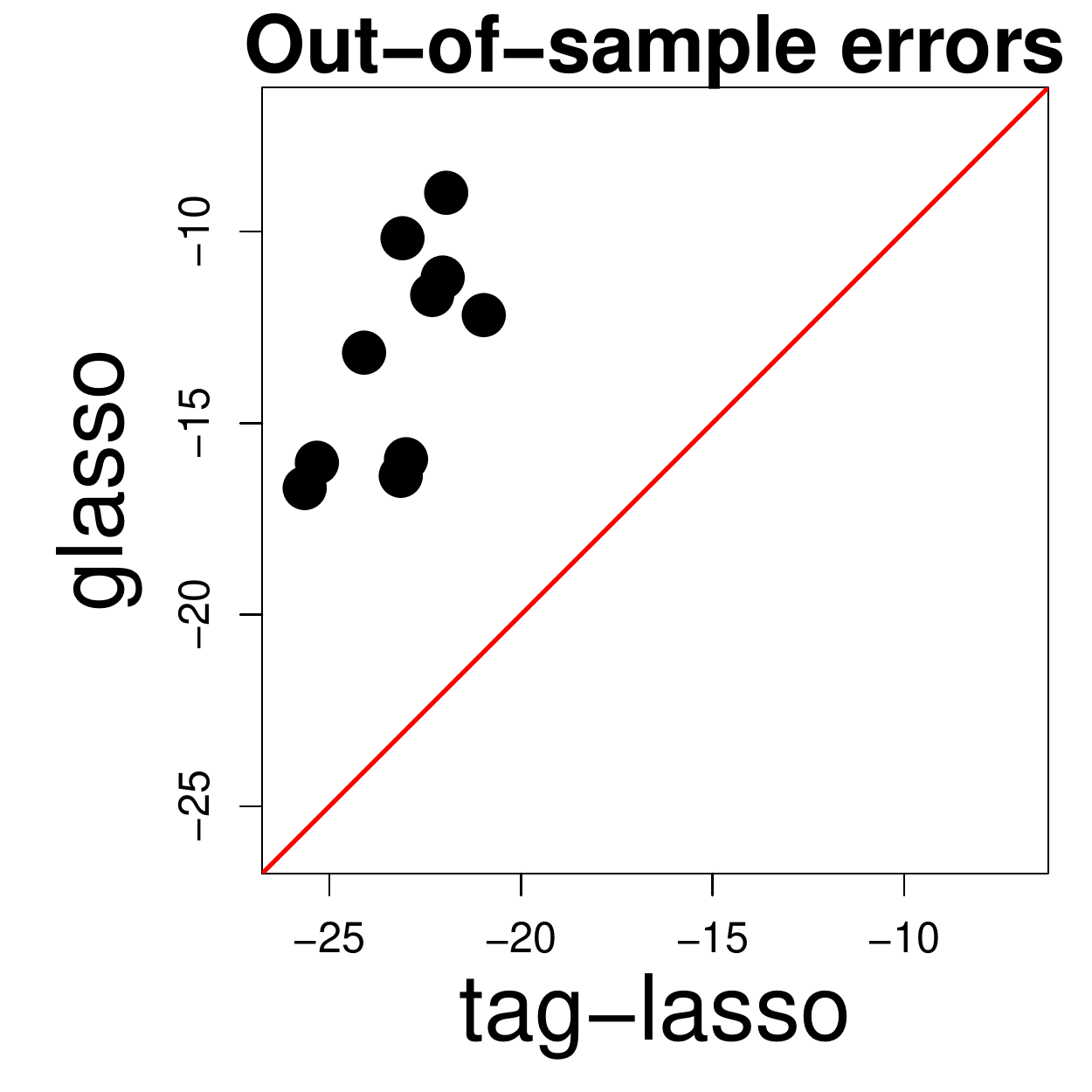}
\includegraphics[width=\textwidth]{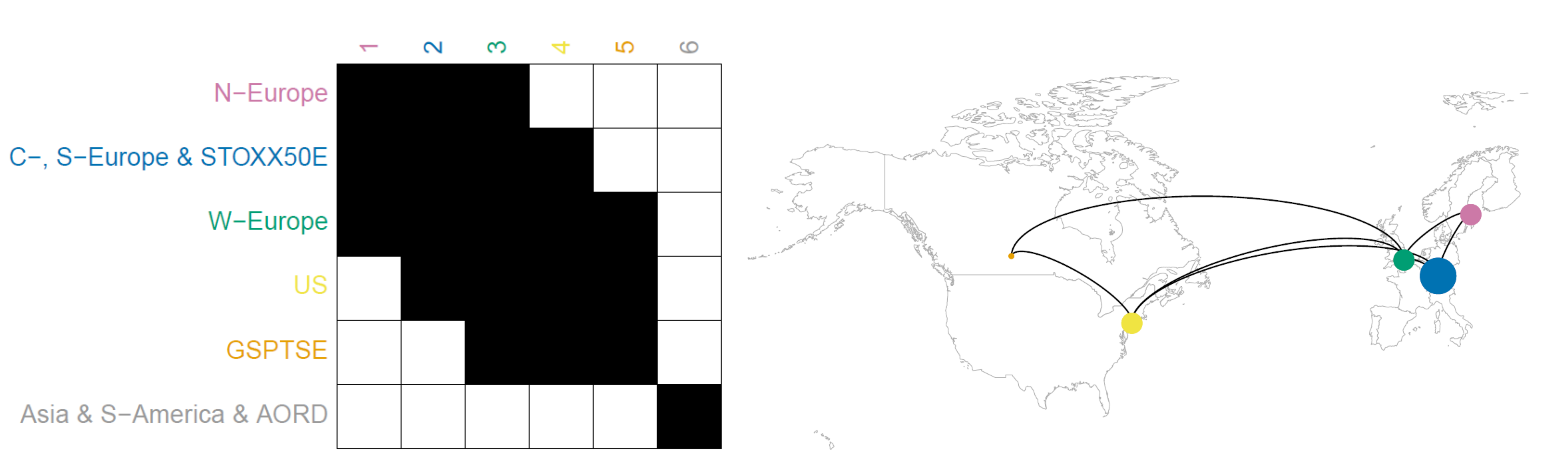}
\caption{Stock market indices data. Top left: Sparsity pattern
  (non-zeros in black) of full $\hat{\boldsymbol\Omega}$ with aggregation
  structure conveyed through row label coloring and column
  numbering. Top right: Test errors across the ten replications (dots)
  for the tag-lasso versus glasso. Bottom: 
Aggregated graph for the $K=6$ nodes obtained with the tag-lasso as an
adjacency matrix (bottom left) and as a network (bottom right) with the size of each node proportional to the number of original variables it aggregates.
\label{stock_market_model}}
\end{figure}

\paragraph{Estimated Graphical Model.}
We fit the tag-lasso estimator, with 5-fold cross validation to select
tuning parameters, to the full data set, with the matrix
${\bf A}$ encoding the tree structure in Figure \ref{stock_market_tree}.
The tag-lasso returns a solution with $K=6$ aggregated blocks; the
sparsity pattern of the full estimated precision matrix is shown in
the top left panel of Figure \ref{stock_market_model}. The coloring of
the row labels and the numbering of columns convey the memberships of
each variable to aggregated blocks (to avoid clutter, only the first
column of each block is labeled).

Dimension reduction mainly occurs through node aggregation, as can be
seen from the aggregated precision matrix in the bottom right panel of
Figure \ref{stock_market_model}. The resulting aggregated graphical
model is  rather dense with only about half of the off-diagonal
entries being non-zero in the estimated aggregated precision matrix, thereby suggesting strong volatility connectedness. 
The solution returned by the tag-lasso estimator consists of one
single-market block (block 5: Canada) and five multi-market blocks, which vary in size.
The Australian, South-America, and all Asian stock markets form one
aggregated block (block 6). 
Note that the tag-lasso has  ``aggregated" these merely because they have the same {\em non}-dependence structure (i.e.\ all of
these markets are estimated to be conditionally independent of each
other and all other markets).
The remaining aggregated nodes concern the US market (block 4) and
three European markets, which are divided into North-Europe (block 1), Central-, South-Europe \& STOXX50E (block 2), and West-Europe (block 3). In the aggregated network, the latter two and the US play a central role as they are the most strongly connected nodes: 
These three nodes are connected to each other,
the US node is additionally connected to Canada, whereas these European nodes are additionally connected with North-Europe.

\paragraph{Out-of-sample Performance.}
We conduct an out-of-sample exercise to compare the tag-lasso estimator to the glasso estimator. 
We take a random $n=203$ observations (80\% of the full data set) to
form a ``training sample" covariance matrix and use the remaining data
to form a ``test sample" covariance matrix ${\bf S}^{\text{test}}$,
and repeat this procedure ten times.
We fit both the tag-lasso and glasso estimator to the training
covariance matrix, with 5-fold cross-validation on the training data
to select tuning parameters. Next, we compute their corresponding
out-of-sample errors on the test data, as in
\eqref{cv-error}. 

The top right panel of Figure \ref{stock_market_model} shows each of
these ten test errors for both the tag-lasso (x-axis) and the glasso
estimator (y-axis). The fact that in all ten replicates the points are
well above the 45-degree line indicates that the tag-lasso estimator
has better estimation error than the glasso.
Tag-lasso has a lower test error than glasso in all ten replicates,
resulting in 
a substantial reduction in glasso's test errors.
This indicates that jointly exploiting edge and node dimension reduction is useful for precision matrix estimation in this context.

\subsection{Microbiome Application}
We next turn to a data set of gut microbial amplicon data in HIV
patients \citep{Rivera-Pinto2018}, where our goal is to estimate an
interpretable graphical model, capturing the interplay between
different taxonomic groups of the microbiome. \citet{bien2020tree}
recently showed that tree-based aggregation in a supervised setting
leads to parsimonious predictive models.  The data set has $n=152$ HIV
patients, and we apply the tag-lasso estimator to all $p=104$
bacterial operational taxonomic units (OTUs) that have non-zero counts
in over half of the samples. We use the taxonomic
tree that arranges the OTUs into natural hierarchical groupings
of taxa: with 17 genera, 11
families, five orders, five classes, three
phyla, and one kingdom (the root node). 
We employ a 
standard data transformation from the field of compositional data analysis (see e.g., \citealp{aitchison1982statistical}) called 
the centered log-ratio (clr) transformation that is commonly used in microbiome graphical modeling \citep{kurtz2015sparse, lo2018pglasso, kurtz2019disentangling}.
After transformation, \citet{kurtz2015sparse} apply the glasso,  \citet{lo2018pglasso} incorporate  phylogenetic information into glasso's optimization problem through weights within the $\ell_1$-penalty, and \citet{kurtz2019disentangling} estimate a latent graphical model which combines sparsity with a low-rank structure. 
We instead, use the tag-lasso to learn a sparse aggregated network
from the clr-transformed microbiome compositions.  While the
clr-transform induces dependence between otherwise independent
components, Proposition 1 in \citet{cao2019large} provides intuition
that as long as the underlying graphical model is sparse and $p$ is large, these induced dependencies may have minimal effect on the covariance matrix.  Future work could more carefully account for the induced dependence, incorporating ideas from \citet{cao2019large} or \citet{kurtz2019disentangling}. 

\begin{figure}
\centering
\includegraphics[width=0.45\textwidth]{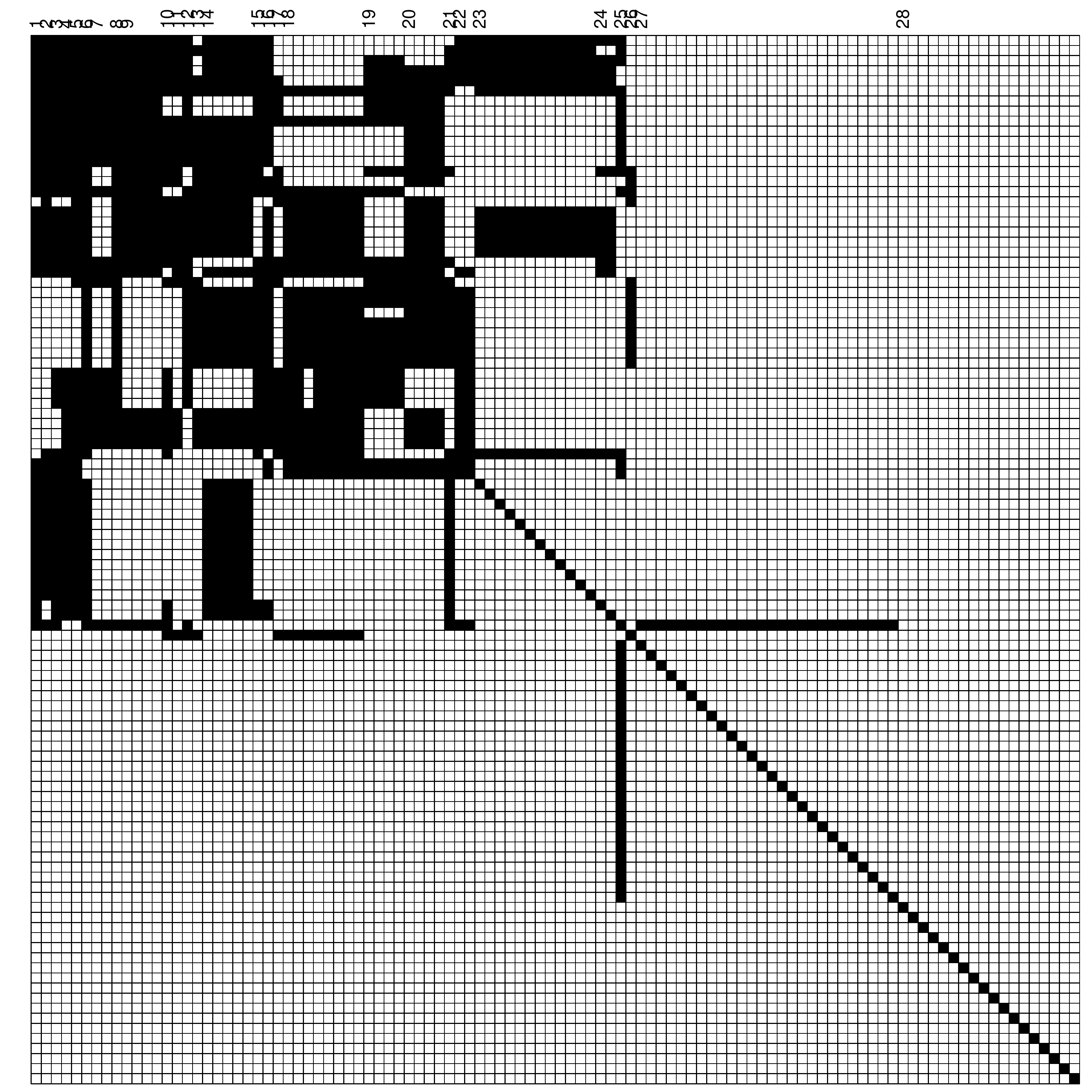}
\includegraphics[width=0.45\textwidth]{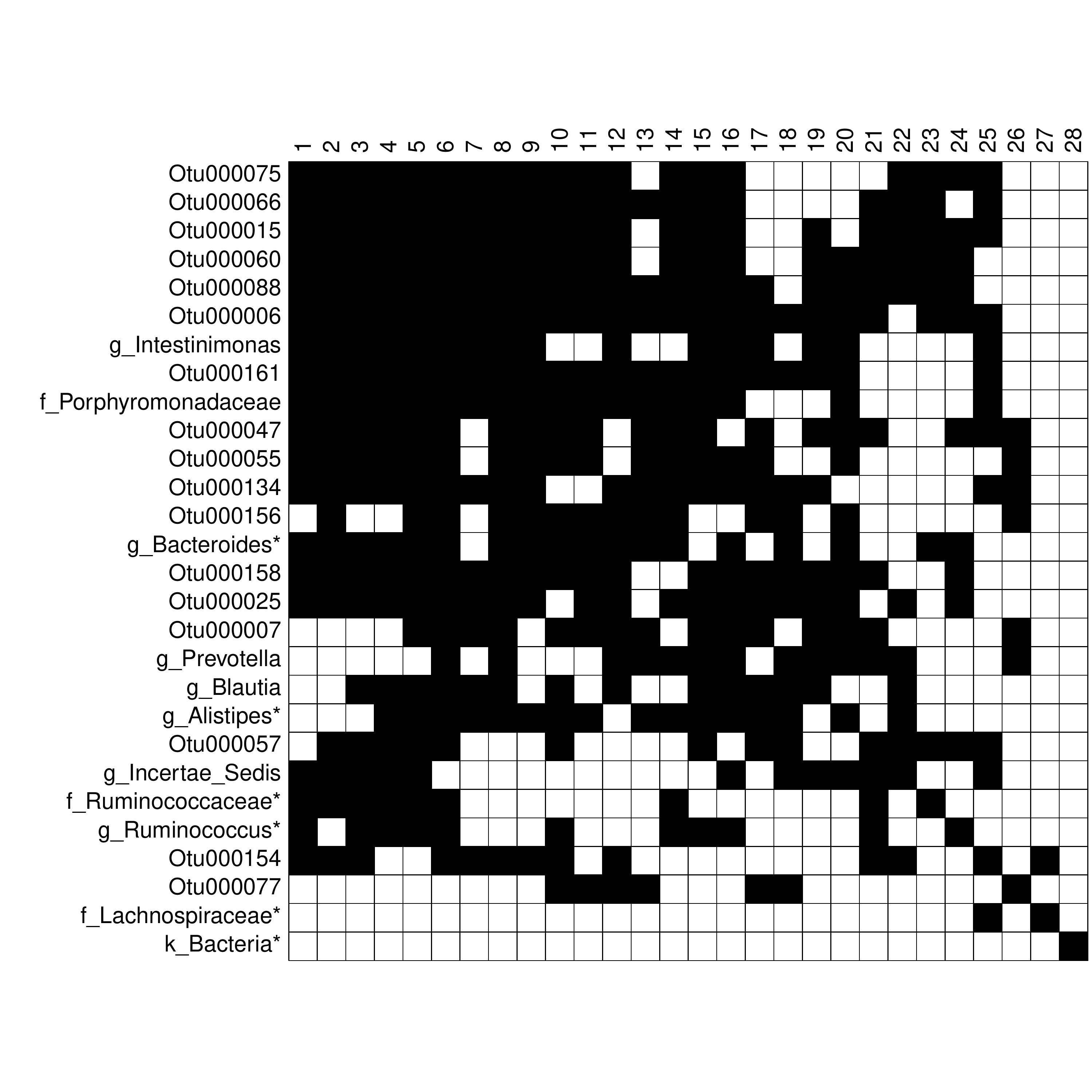}

\includegraphics[width=0.45\textwidth]{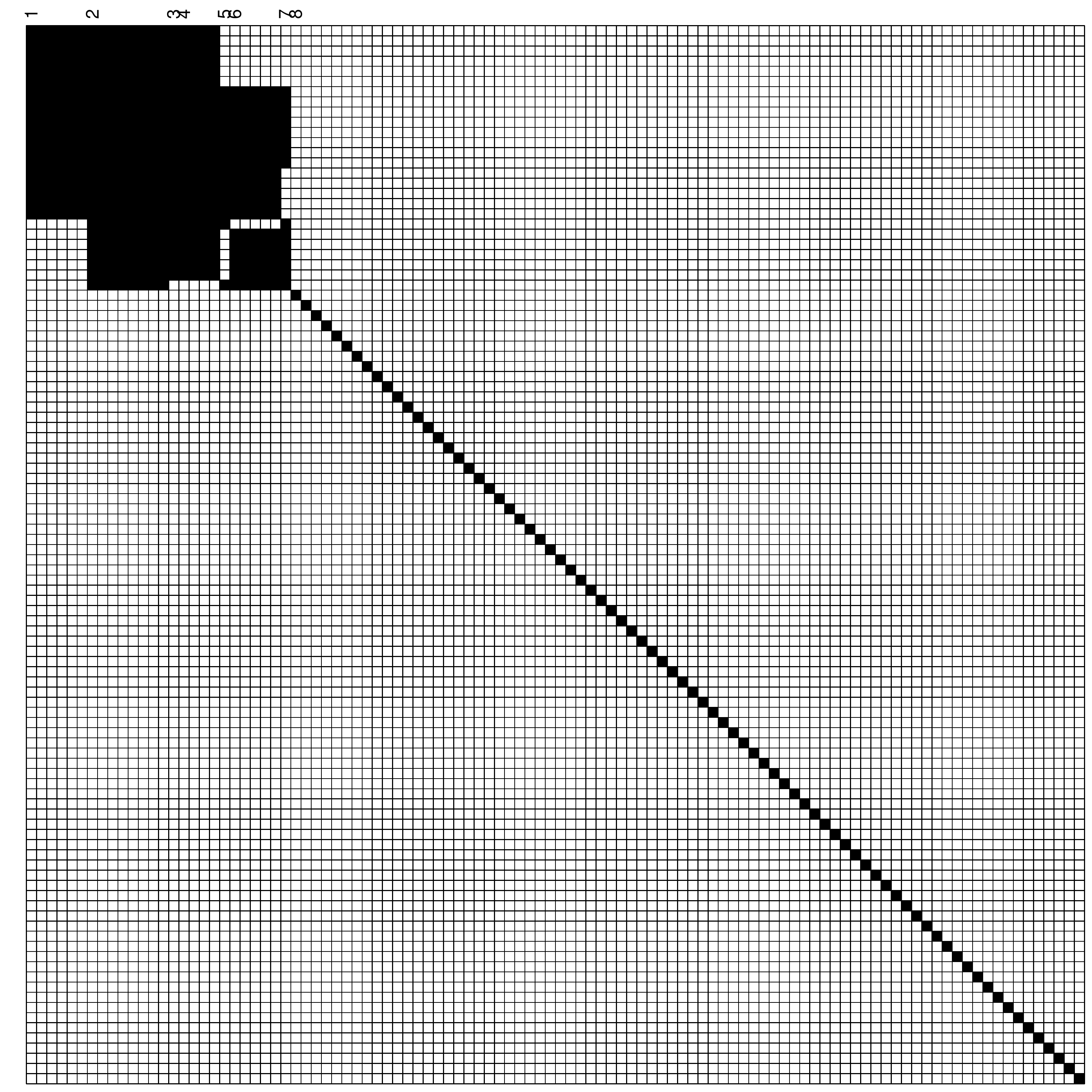}
\includegraphics[width=0.45\textwidth]{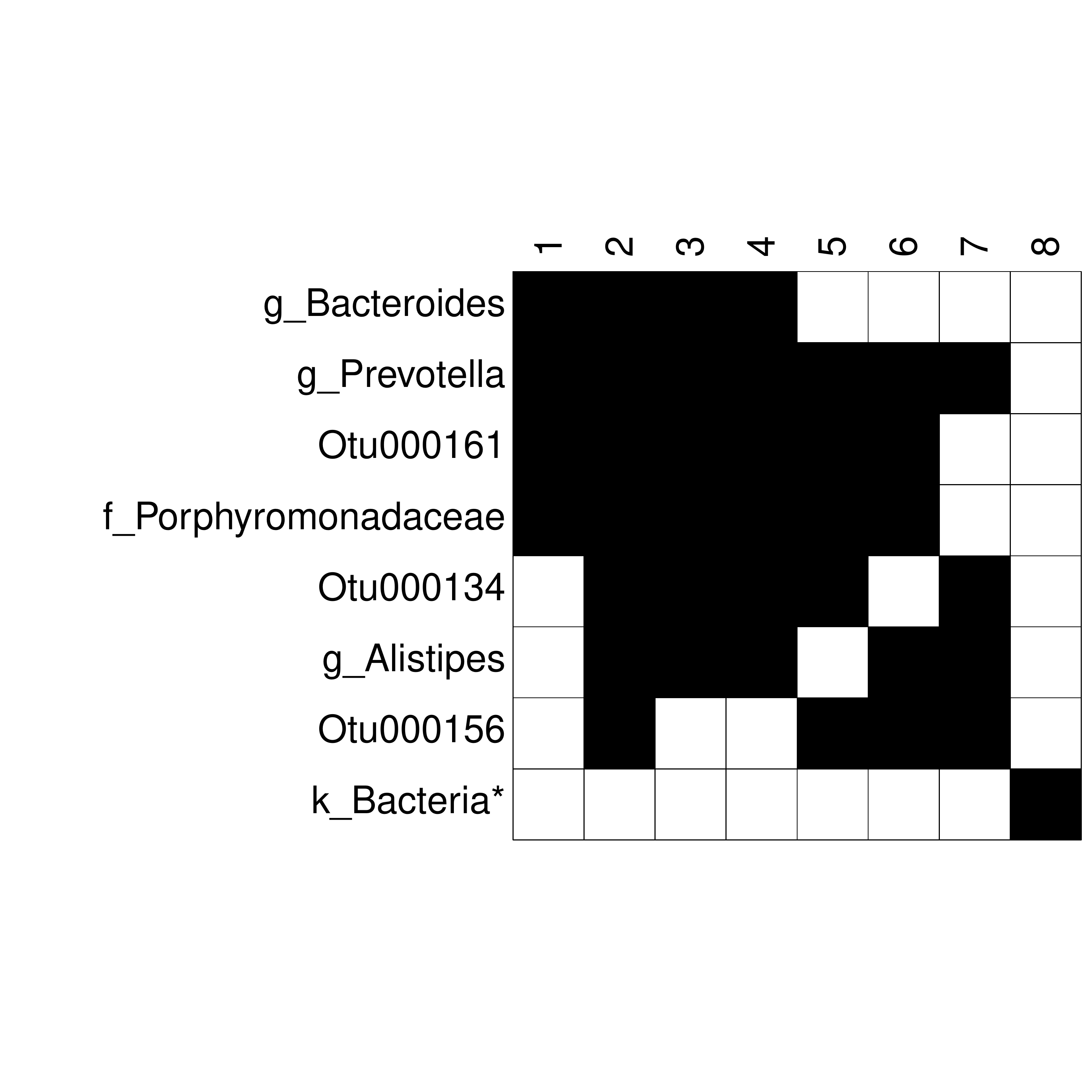}
\caption{Microbiome data. Full precision matrix (left) and aggregated
  precision matrix (right) estimated by the tag-lasso with an
  unconstrained five-fold cross-validation (top) and with a
  cross-validation subject to the constraint that there are at most
  ten blocks (bottom).
\label{application_OTU}}
\end{figure}

\paragraph{Estimated Graphical Model.}
We fit the tag-lasso to the full data set and use 5-fold cross-validation to select the tuning parameters. 
The tag-lasso estimator provides a sparse aggregated graphical model
with $K=28$ aggregated blocks (a substantial reduction in nodes from
the original $p=104$ OTUs). The top panel
of Figure \ref{application_OTU} shows the sparsity pattern of the
$p\times p$ estimated precision matrix (top left) and of the $K \times
K$ estimated aggregated precision matrix (top right).
A notable feature of the tag-lasso solution is that it returns a wide
range of aggregation levels: The aggregated network consists of 17 OTUs, 7 nodes aggregated to the genus level (these nodes start with ``g\_"), 3 to the family level (these nodes start with``f\_"), and 1 node to the kingdom level (this node starts with ``k\_"). 
Some aggregated nodes, such as the ``g\_Blautia" node (block 19),
contain all OTUs within their taxa; some other
aggregated nodes, indicated with an asterisk like the ``k\_Bacteria*" node (block 28), have some of their
OTUs missing. 
This latter ``block" consists of 18 OTUs from across the
phylogenetic tree that are estimated to be conditionally independent with all other OTUs in the data set.

\begin{figure}
\centering
\includegraphics[width=0.32\textwidth]{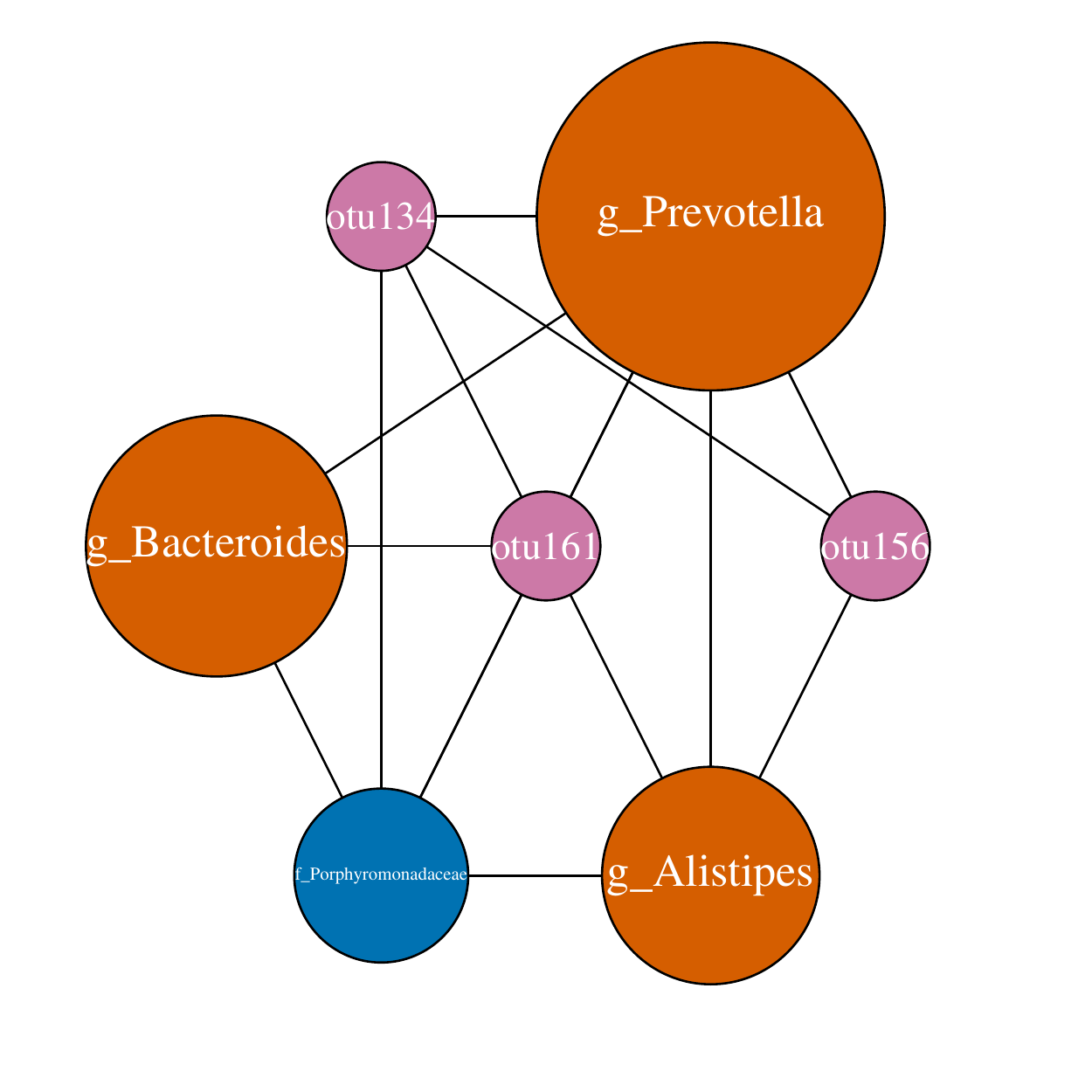}
\hspace{-0.5cm}
\includegraphics[width=0.32\textwidth]{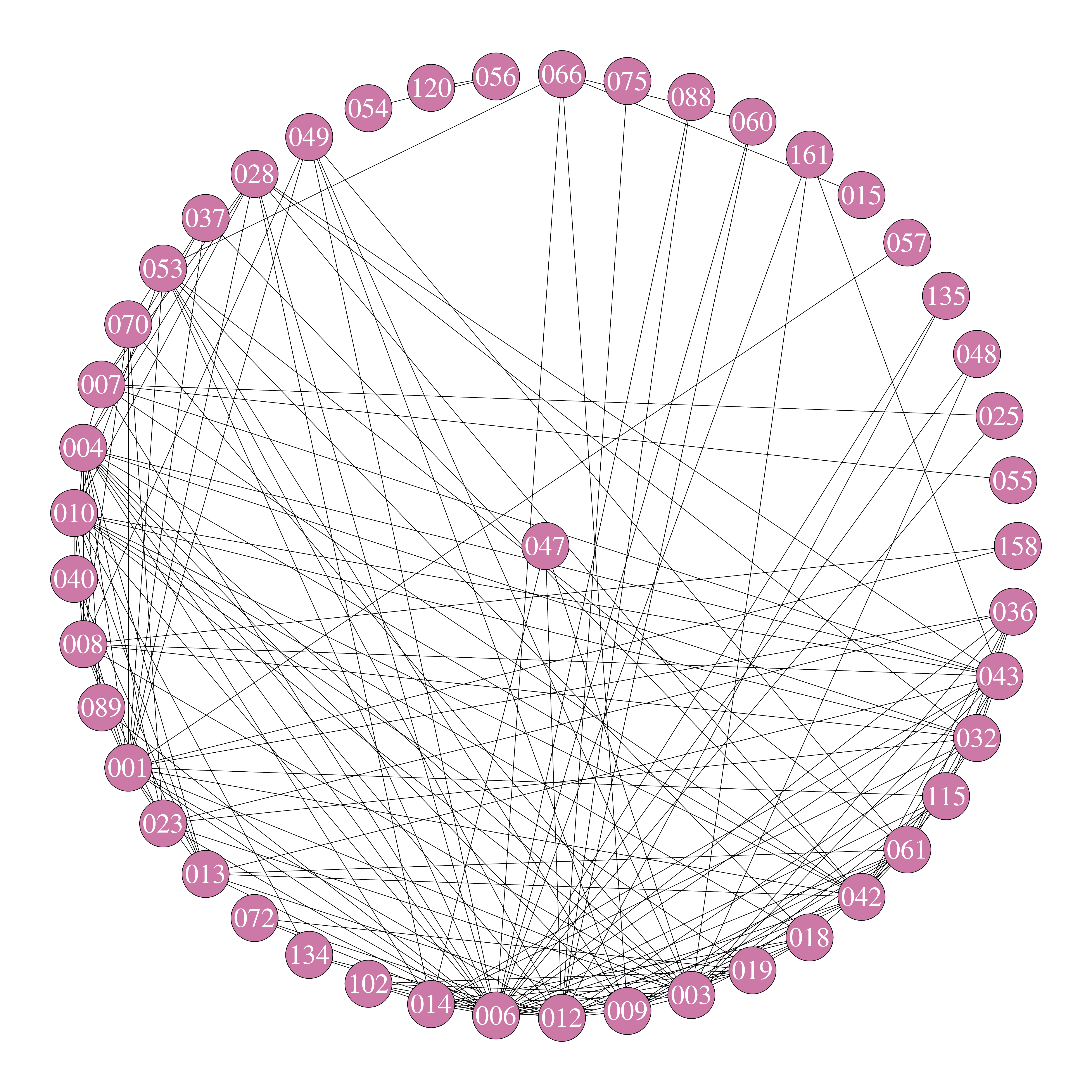}
\includegraphics[width=0.35\textwidth]{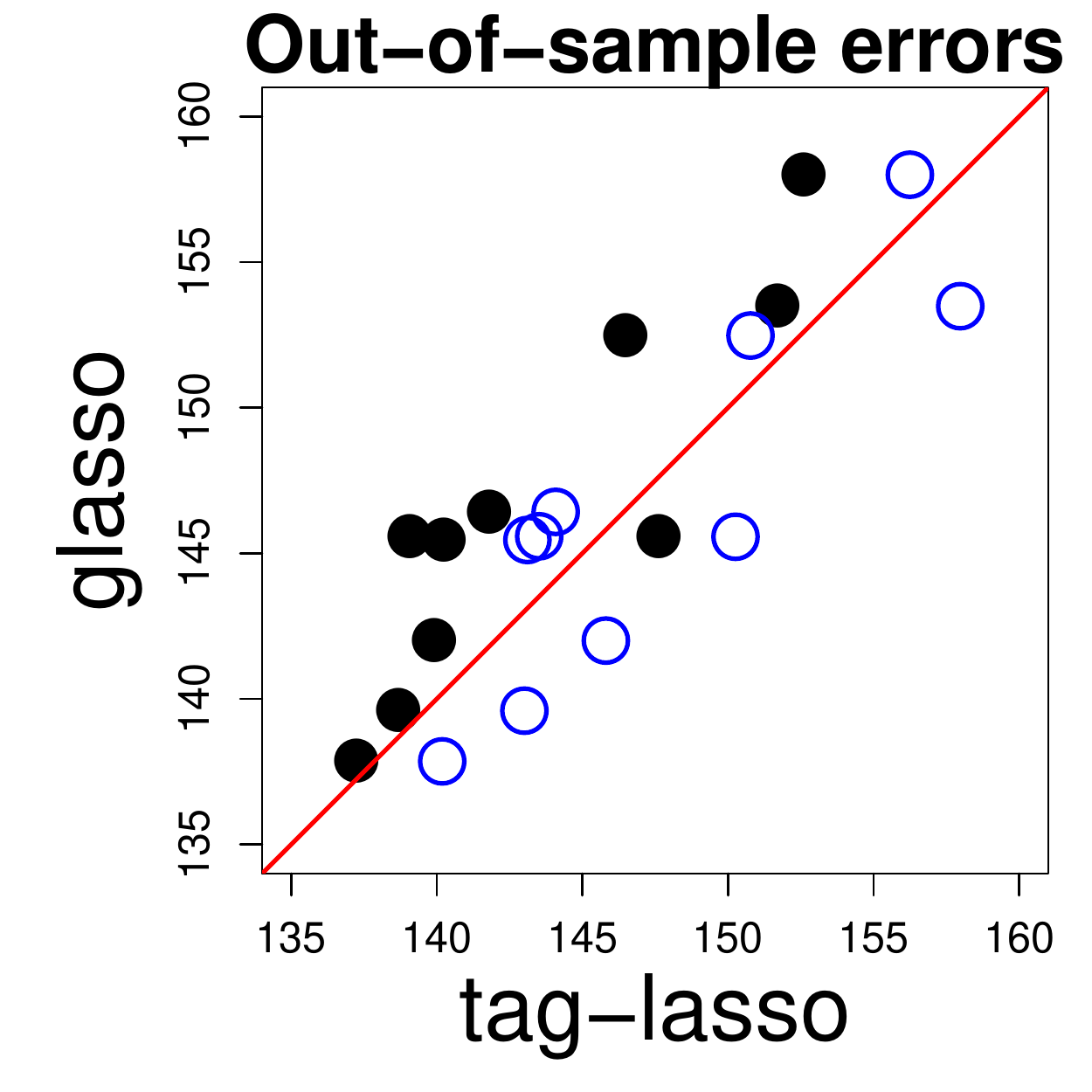}
\caption{Microbiome data. 
Left: Aggregated network estimated by the constrained CV version of the tag-lasso. The colour of the nodes is based on their level of aggregation (OTU: pink, genus: orange, family: blue); their width is proportional to the number of OTUs they aggregate.
Middle: Network estimated by the glasso.
Right: Test errors across the ten replications for the unconstrained (solid black) and constrained (unfilled blue) CV version of the tag-lasso versus the glasso. 
}
\label{OTU_network_oos}
\end{figure}

While the tag-lasso determines the aggregation level in a data-driven way through cross validation, practitioners or researchers may also sometimes wish to restrict the number of blocks $K$ to a pre-determined level when such prior knowledge is available or if this is desirable for interpretability. As an illustration, we consider a {\em constrained cross-validation} scheme in which we restrict the number of blocks $K$ to maximally ten and select the sparsity parameters with the best cross validated error among those solutions with $K\leq 10$.
The bottom panel of Figure \ref{application_OTU} shows the sparsity pattern of the full and aggregated precision matrices estimated by this constrained version of the tag-lasso. 

The resulting network consists of $K=8$ aggregated nodes. The
``k\_Bacteria*" node now aggregates 78 OTUs that are estimated to be
conditionally independent with each other and all others. The
interactions among the remaining nodes are shown in the left panel  of
Figure \ref{OTU_network_oos}, which consists of three OTUs (OTU134,
OTU156, and OTU161, in pink), three genera (Prevotella, Bacteroides,
and Alistipes in orange) and one family (Porphyromonadaceae in blue).
The resulting network is much simpler than the one estimated by the
glasso, shown in the middle panel of Figure \ref{OTU_network_oos}. The
glasso finds 58 OTUs to be conditionally independent with all others,
but the interactions among the remaining 46 OTUs are much more
difficult to interpret.  The glasso is limited to working at the
OTU-level, which prevents it from providing insights about
interactions that span different levels of the taxonomy.

\paragraph{Out-of-sample Performance.} We conduct the same out-of-sample exercise as described in Section \ref{finance}. 
The right panel of Figure \ref{OTU_network_oos}
presents the ten test errors (black dots) for the unconstrained CV tag-lasso and
glasso. In all but one case, the tag-lasso leads to a better fit than the glasso, suggesting that it is better suited for modeling the conditional dependencies among the OTUs.
The unfilled blue dots show the same but for the constrained CV
tag-lasso. In all ten cases, it underperforms the unconstrained CV
tag-lasso (see shift to the right on the horizontal axis); however,
its performance is on a par with the glasso, with test errors close to
the 45 degree line.  Thus, there does not appear to be a cost in
out-of-sample-performance to the
interpretability gains of the constrained tag-lasso over the
glasso.

\section{Conclusion} \label{conclusion}
Detecting conditional dependencies between variables, as represented
in a graphical model, forms a cornerstone of multivariate data
analysis.
However, graphical models, characterized by a set of nodes and edges, can
quickly explode in dimensionality due to ever-increasing
fine-grained levels of resolution at which data are measured.  In
many applications, a tree is available that organizes the measured
variables into various meaningful levels of resolution.
In this work, we introduce the tag-lasso, a novel estimation procedure for graphical
models that curbs this curse of dimensionality through joint node and
edge dimension reduction by leveraging this tree as side
information. Node dimension reduction is achieved by a penalty that allows nodes to
be aggregated according to the tree structure; edge dimension reduction is achieved through a standard sparsity-inducing penalty. 
As such, the tag-lasso generalizes the popular glasso approach to sparse graphical modelling. An \verb|R| package called \verb|taglasso| implements the proposed method and is available on the GitHub page of the first author.

\paragraph{Acknowledgments} We thank Christian M{\"u}ller for useful discussions.  Jacob
Bien was supported in part by NSF CAREER Award
DMS-1653017 and NIH Grant R01GM123993.
\bibliographystyle{asa}
\bibliography{refs}

\newpage
\appendix 
\section*{Appendices}
\section{Proof of Proposition \ref{omega_p_to_K}} \label{app_proof_prop}

\begin{proof} 
First, note that ${\bf \widetilde{X}}$ follows a $K$-dimensional multivariate normal distribution with mean zero and covariance matrix ${\bf M}^\top ({\bf D} + {\bf M}{\bf C}{\bf M}^\top )^{-1} {\bf M}$.
Next, we re-write this covariance matrix by two successive applications of 
equation (23) in \cite{henderson1981deriving}:
\color{black}
\begin{eqnarray}
{\bf M}^\top ({\bf D} + {\bf M}{\bf C}{\bf M}^\top)^{-1} {\bf M} & = & 
{\bf M^\top}{\bf D}^{-1}{\bf M} - {\bf M^\top}{\bf D}^{-1}{\bf M}({\bf I} + {\bf C}{\bf M^\top}{\bf D}^{-1}{\bf M})^{-1}{\bf C}{\bf M^\top}{\bf D}^{-1}{\bf M} \nonumber \\
& = & \left([{\bf M^\top}{\bf D}^{-1}{\bf M}]^{-1} + {\bf C}\right)^{-1}. \nonumber 
\end{eqnarray}
\color{black}
Hence, the precision matrix of ${\bf \widetilde{X}}$ is given by $({\bf M^\top}{\bf D}^{-1}{\bf M})^{-1} + {\bf C}$. Now since $({\bf M^\top}{\bf D}^{-1}{\bf M})^{-1}$ is diagonal, 
$c_{ij} = 0 \Leftrightarrow \widetilde{X}_i \ \bot \  \widetilde{X}_j | {\bf \widetilde{X}}_{-\{i,j\}}$, for any $i, j = 1, \ldots, K$ and with  ${\bf \widetilde{X}}_{-\{i,j\}}$ containing all aggregated variables expect for aggregate $i$ and $j$.
\end{proof}

\section{Details of the LA-ADMM Algorithm} \label{Appendix_ADMM}
The augmented Lagrangian of \eqref{ADMMglobal} is given by 
\begin{footnotesize}
\begin{eqnarray}
&& -\text{logdet}(\boldsymbol\Omega^{(1)}) + \text{tr}({\bf S}\boldsymbol{\Omega}^{(1)}) + 
1_{\infty}\{\boldsymbol{\Omega}^{(1)}={\boldsymbol{\Omega}^{(1)}}^\top,  \boldsymbol \Omega^{(1)} \succ {\bf 0}\} + 
\langle {\bf U}^{(1)}, {\boldsymbol \Omega}^{(1)} - {\boldsymbol \Omega} \rangle  + \dfrac{\rho}{2} \|{\boldsymbol \Omega}^{(1)} - {\boldsymbol \Omega} \|^2_F  \nonumber \\
& + & \lambda_1 \|\boldsymbol{\Gamma}^{(1)}_{-r}\|_{2,1} +  1_{\infty}\{\boldsymbol\gamma_r^{(1)} = \gamma^{(1)} {\bf 1}_p\} \  + 
\langle {\bf U}^{(4)}, {\boldsymbol \Gamma}^{(1)} -  {\boldsymbol \Gamma} \rangle  + \dfrac{\rho}{2} \|{\boldsymbol \Gamma}^{(1)} - {\boldsymbol \Gamma}\|^2_F \nonumber \\
& + & 1_{\infty}\{\boldsymbol \Omega^{(2)} = {\bf A}\boldsymbol{\Gamma}^{(2)}+ {\bf D}, {\bf D} \ \text{diag.}, D_{jj} \geq 0\} + 
\langle {\bf U}^{(2)}, {\boldsymbol \Omega}^{(2)} - {\boldsymbol \Omega} \rangle  + \dfrac{\rho}{2} \|{\boldsymbol \Omega}^{(2)} - {\boldsymbol \Omega} \|^2_F 
+ 
\langle {\bf U}^{(5)}, {\boldsymbol \Gamma}^{(2)} -  {\boldsymbol \Gamma} \rangle  + \dfrac{\rho}{2} \|{\boldsymbol \Gamma}^{(2)} - {\boldsymbol \Gamma}\|^2_F
\nonumber \\
& + & \lambda_2 \|\boldsymbol \Omega^{-\text{diag}(3)}\|_1  + 
\langle {\bf U}^{(3)}, {\boldsymbol \Omega}^{(3)} - {\boldsymbol \Omega} \rangle  + \dfrac{\rho}{2} \|{\boldsymbol \Omega}^{(2)} - {\boldsymbol \Omega} \|^2_F,  \label{Lagrangian}
\end{eqnarray}
\end{footnotesize}
where ${\bf U}^{(i)}$ (for $i=1, \ldots, 5)$ are the dual variables, and $\rho$ is a penalty parameter.
Note that equation \eqref{Lagrangian} is of the same form as Equation (3.1) in \cite{Boyd11} and thus involves iterating three basic steps:
(i) minimization with respect to  
$(\boldsymbol{\Omega}^{(1)}, \boldsymbol{\Omega}^{(2)}, \boldsymbol{\Omega}^{(3)}, \boldsymbol{\Gamma}^{(1)}, \boldsymbol{\Gamma}^{(2)}, {\bf D})$,
(ii) minimization with respect to 
$(\boldsymbol{\Omega}, \boldsymbol{\Gamma})$, and
(iii) update of $({\bf U}^{(1)}, \ldots, {\bf U}^{(5)})$. 

Step (i)
decouples into four independent problems, whose solutions are worked
out in Sections \ref{solveom1}-\ref{om3}.
Step (ii) involves the minimization of a differentiable function of  ${\bf \Omega}$ and ${\bf \Gamma}$ and boils down to the calculation of simple averages, as shown in Section \ref{global}. 
Step (iii)'s update of the dual variables is provided in \ref{dual}.  

Algorithms \ref{ADMM_inner}-\ref{LA-ADMM} then provide an overview of the LA-ADMM algorithm to solve problem \eqref{ADMMglobal}. We use the LA-ADMM algorithm with $\rho_1 = 0.01, \  \texttt{T}_\text{stages} = 10, \ \texttt{maxit} = 100$.

\begin{algorithm}[H]
	\caption{ADMM}
	 \color{black}\label{ADMM_inner}
	\begin{description}
		\item[Input:] ${\bf S}, {\bf A}, p, |\mathcal{T}|,  \lambda_1, \lambda_2, \rho, \texttt{maxit}, \boldsymbol{\Omega}_0, \boldsymbol{\Gamma}_0.$
		
		\item[Initialization:] Set
		\begin{itemize}[label={}]
			
			\item $\widehat{{\boldsymbol \Omega}}^{(i)}_{0} \leftarrow 
			\widehat{{\bf U}}^{(i)}_{0} \leftarrow \boldsymbol{\Omega}_0\ \hspace{0.4cm} \text{for} \ i=1, \ldots, 3$
			\item $ \widehat{{\boldsymbol \Gamma}}^{(j)}_{0} \leftarrow \widehat{{\bf U}}^{(j+3)}_{0} \leftarrow \boldsymbol{\Gamma}_0\ \text{for} \ j=1, \ldots, 2$
			\item $ k \leftarrow 0$
		\end{itemize}
	\item[for] $k \leq {\tt maxit}$ \textbf{do}
	\begin{itemize}[label={}]
	\vspace{-0.4cm}
		\item $k \leftarrow k + 1$
		\item $\widehat{{\boldsymbol \Omega}}_{k}^{(1)} \leftarrow {\bf Q} {\bar {\boldsymbol \Omega}}_{k-1}{\bf Q}^\top$, see equation \eqref{Omega1}.
		\item $\widehat{\Omega}^{(3)}_{k, ij} \leftarrow  S({\widehat\Omega}_{k-1, ij} - { {\widehat U}}_{k-1, ij}^{(3)}/\rho, \ \lambda_2/\rho), \forall \ i,j=1,\ldots, p$, see equation \eqref{Omega3}.
		\item $\boldsymbol{\widehat{\Gamma}}^{(1)}_{k, j} \leftarrow S_G(\boldsymbol{\widehat\Gamma}_{k-1, j} - { \bf{\widehat U}}_{k-1, j}^{(4)}/\rho, \ \lambda_1/\rho), \forall j=1,\ldots,|\mathcal{T}|\backslash\{r\}$, see equation \eqref{Gamma1}.
		\item $\boldsymbol{\widehat{\Gamma}}^{(1)}_{k, r} \leftarrow \widehat{\gamma}_{k-1} {\bf 1}_p$, see equation \eqref{Gamma1}.
		\item $\text{diag}({\bf {\widehat D}}_{k}) \leftarrow   \text{diag}({\bf C}^\top {\bf C})^{-1}\text{diag}({\bf B}^\top {\bf C})_+$, see equation \eqref{equationD}.
		\item $\boldsymbol {\widehat\Gamma}^{(2)}_{k} \leftarrow   ({\bf A}^\top{\bf A} + {\bf I}_{|\mathcal{T}|})^{-1}({\bf A}^\top : {\bf I}_{|\mathcal{T}|})({\bf \widetilde M} - {\bf {\widetilde D}}_{k})$, see equation \eqref{Gamma2}.
		\item $\widehat{\boldsymbol \Omega}^{(2)}_{k} = {\bf A}\widehat{\boldsymbol \Gamma}^{(2)}_{k} + {\bf {\widehat D}}_{k} $, see equation \eqref{Omega2}
		\smallskip
		\item $\widehat{\boldsymbol \Omega}_{k} \leftarrow (\widehat{\boldsymbol \Omega}^{(1)}_{k} + \widehat{\boldsymbol \Omega}^{(2)}_{k} + \widehat{\boldsymbol \Omega}^{(3)}_{k})/3 $
		\item $\widehat{\boldsymbol \Gamma}_{k} \leftarrow (\widehat{\boldsymbol \Gamma}^{(1)}_{k} + \widehat{\boldsymbol \Gamma}^{(2)}_{k})/2 $
		\smallskip
		\item $\widehat{\bf U}^{(i)}_{k} \leftarrow   \widehat{\bf U}^{(i)}_{k-1} + \rho \left ( \widehat{{\boldsymbol \Omega}}_{k}^{(i)} - \widehat{{\boldsymbol \Omega}}_{k} \right), \ \text{for} \ i=1, \ldots, 3$ 
		\item $\widehat{\bf U}^{(j+3)}_{k} \leftarrow  \widehat{\bf U}^{(j+3)}_{k-1} + \rho \left ( \widehat{{\boldsymbol \Gamma}}_{k}^{(j)} - \widehat{{\boldsymbol \Gamma}}_{k} \right), \ \text{for} \ j=1, \ldots, 2$
	\end{itemize}	
	\item[\textbf{end for}]
	\item[Output:] $\widehat{{\boldsymbol \Omega}}_{\texttt{maxit}}, \ \widehat{{\boldsymbol \Gamma}}_{\texttt{maxit}},  \ \widehat{{\bf D}}_{\texttt{maxit}}$
	\end{description}
\end{algorithm}

\begin{algorithm}[H]
	\caption{LA-ADMM} \label{LA-ADMM}
	\color{black}
	\begin{description}
		\item[Input:] ${\bf S}, {\bf A}, p, |\mathcal{T}|, \lambda_1, \lambda_2, \rho_1, \texttt{maxit}, \texttt{T}_{\text{stages}}$
				\item[Initialization:] Set
		\begin{itemize}[label={}]
			\item $\widehat{{\boldsymbol \Omega}}_0 \leftarrow {\bf 0}$; $\widehat{{\boldsymbol \Gamma}}_0 \leftarrow {\bf 0}$
			\item $ t \leftarrow 0$

		\end{itemize}
		\item[for] $t \leq \texttt{T}_{\text{stages}}$ \textbf{do}
		\begin{itemize}[label={}]
			\item $t \leftarrow t + 1$
			\item $(\boldsymbol{\widehat{\Omega}}_t, \boldsymbol{\widehat{\Gamma}}_t, \widehat{{\bf D}}_t) \leftarrow \text{ADMM}({\bf S}, {\bf A}, p, |\mathcal{T}|, \lambda_1,  \lambda_2, \rho_{t}, \texttt{maxit}, \widehat{{\boldsymbol \Omega}}_{t-1}, \widehat{{\boldsymbol \Gamma}}_{t-1}) $
			\item $\rho_{t+1} \leftarrow 2\rho_{t}  $ 
		\end{itemize}	
		\item[\textbf{end for}]
		\item[Output:] $\widehat{{\boldsymbol \Omega}}_{\texttt{T}_{\text{stages}}}, \ \widehat{{\boldsymbol \Gamma}}_{\texttt{T}_{\text{stages}}}, \ \widehat{{\bf D}}_{\texttt{T}_{\text{stages}}}$
	\end{description}
\end{algorithm}

\subsection{Solving for $\boldsymbol \Omega^{(1)}$} \label{solveom1}
Minimizing the augmented Lagrangian with respect to $\boldsymbol \Omega^{(1)}$ gives
\begin{small}
\begin{eqnarray}
\widehat{\boldsymbol \Omega}^{(1)}_{k+1} & := & \underset{\boldsymbol \Omega^{(1)}}{\operatorname{argmin}} \{- \text{logdet}(\boldsymbol\Omega^{(1)}) + \text{tr}({\bf S}\boldsymbol{\Omega}^{(1)}) + \langle {\bf U}^{(1)}, {\boldsymbol \Omega}^{(1)} - {\boldsymbol \Omega} \rangle + \dfrac{\rho}{2} \|{\boldsymbol \Omega}^{(1)} - {\boldsymbol \Omega} \|^2_F  \ \  \text{s.t.} \  \  \boldsymbol{\Omega}^{(1)}={\boldsymbol{\Omega}^{(1)}}^\top, 
\boldsymbol \Omega^{(1)} \succ {\bf 0}
\}  \nonumber \\
& = & \underset{\boldsymbol \Omega^{(1)}}{\operatorname{argmin}} \{- \text{logdet}(\boldsymbol\Omega^{(1)}) + \text{tr}({\bf S}\boldsymbol{\Omega}^{(1)}) + \dfrac{\rho}{2} \| \boldsymbol \Omega^{(1)} - (\boldsymbol {\widehat\Omega}_k - {\bf {\widehat U}}_k^{(1)}/\rho)\|^2_F  \ \  \text{s.t.} \  \  \boldsymbol{\Omega}^{(1)}={\boldsymbol{\Omega}^{(1)}}^\top, 
\boldsymbol \Omega^{(1)} \succ {\bf 0}
\} \label{ADMMOmega1} \nonumber
\end{eqnarray}
\end{small}
The solution should satisfy the first order optimality condition
\begin{equation}
\rho \boldsymbol {\widehat\Omega}^{(1)}_{k+1} - {\boldsymbol{\widehat\Omega}_{k+1}^{(1)}}^{-1} =  \rho \boldsymbol {\widehat\Omega}_k - {\bf {\widehat U}}_k^{(1)}- {\bf S}.
\label{Omega1FOC}
\end{equation}
This means that the eigenvectors of $\boldsymbol {\widehat\Omega}^{(1)}_{k+1}$ are the same as the eigenvectors of $\rho \boldsymbol {\widehat\Omega}_k - {\bf {\widehat U}}_k^{(1)}- {\bf S}$ and that the eigenvalues of $\boldsymbol {\widehat\Omega}^{(1)}_{k+1}$ are a simple function of the eigenvalues of $\rho \boldsymbol {\widehat\Omega}_k - {\bf {\widehat U}}_k^{(1)}- {\bf S}$.
Consider the orthogonal eigenvalue decomposition of right hand side:
$$ \rho \boldsymbol {\widehat\Omega}_k - {\bf {\widehat U}}_k^{(1)}- {\bf S} = {\bf Q} {\boldsymbol \Lambda} {\bf Q}^\top,$$ 
where ${\boldsymbol \Lambda}= {\text{\bf diag}}(\delta_1, \ldots, \delta_p)$ and ${\bf Q}{\bf Q}^\top = {\bf Q}^\top{\bf Q} = {\bf I}$.
Multiply \eqref{Omega1FOC} by ${\bf Q}^\top$ on the left and ${\bf Q}$ on the right
$$ \rho {\boldsymbol {\bar\Omega}}^{(1)}_{k+1} - {\boldsymbol{\bar \Omega}_{k+1}^{(1)}}^{-1} = {\boldsymbol \Lambda}, \ \text{with} \ {\boldsymbol {\bar\Omega}}^{(1)}_{k+1} = {\bf Q}^{\top} {\boldsymbol {\widehat\Omega}}_{k+1}^{(1)}{\bf Q}.$$
Then
\begin{equation}
{\boldsymbol {\widehat\Omega}}_{k+1}^{(1)} = {\bf Q}{\boldsymbol {\bar\Omega}}_{k+1}^{(1)} {\bf Q}^\top, \ \text{with} \ {{\bar\Omega}}^{(1)}_{k+1,jj} = \dfrac{\delta_j + \sqrt{\delta_j^2 + 4 \rho}}{2\rho}. \label{Omega1}
\end{equation}

\subsection{Solving for $\boldsymbol \Gamma^{(1)}$}
Minimizing the augmented Lagrangian with respect to $\boldsymbol \Gamma^{(1)}$ gives
\begin{equation}
\widehat{\boldsymbol \Gamma}^{(1)}_{k+1} :=  \underset{\boldsymbol \Gamma^{(1)}}{\operatorname{argmin}} \{\dfrac{\rho}{2} \| \boldsymbol \Gamma^{(1)} - (\boldsymbol {\widehat\Gamma}_k - {\bf {\widehat U}}_k^{(4)}/\rho)\|^2_F + \lambda_1 \|\boldsymbol{\Gamma}^{(1)}_{-r}\|_{2,1} \ \text{s.t.} \ \boldsymbol\gamma_r^{(1)} = \gamma^{(1)} {\bf 1}_p
\}  \label{ADMMGamma1} \nonumber 
\end{equation}
The solution is  groupwise soft-thresholding:
\begin{equation}
\boldsymbol{\widehat{\Gamma}}^{(1)}_{k+1, j} = \begin{cases}
S_G(\boldsymbol{\widehat\Gamma}_{k, j} - { \bf{\widehat U}}_{k, j}^{(4)}/\rho, \lambda_1/\rho), & \text{if} \ j=1,\ldots,|\mathcal{T}|\backslash\{r\} \\
\widehat{\gamma}_k {\bf 1}_p, & \text{if} \ j=r.
\end{cases} \label{Gamma1}
\end{equation}
with the group-wise soft-thresholding operator  
$S_G({\boldsymbol \gamma}, \lambda) = \text{max}(1 - \lambda/\|{\boldsymbol \gamma}\|_2, 0){\boldsymbol \gamma}$ applied to $\boldsymbol\gamma \in \mathds{R}^p$, and $\widehat{\gamma}_k$ is equal to the average of the $p$-dimensional vector $\boldsymbol{\widehat\Gamma}_{k, r} - { \bf{\widehat U}}_{k, r}^{(4)}/\rho$.
Note that in this Appendix  we use the capitalized $\boldsymbol\Gamma_j$ notation to index the $j^{\text{th}}$ row of the matrix $\boldsymbol\Gamma$ whereas we use lowercase $\boldsymbol\gamma_u$ when indexing a node $u$ based on the tree structure in Section \ref{node_aggregation} of the main paper.

\subsection{Solving for $\boldsymbol \Omega^{(2)}, \boldsymbol \Gamma^{(2)}, {\bf D}$} \label{solvedog}
Minimizing the augmented Lagrangian with respect to $\boldsymbol \Omega^{(2)}, \boldsymbol \Gamma^{(2)}, {\bf D}$ gives
\begin{multline}
(\widehat{\boldsymbol \Omega}^{(2)}_{k+1}, \widehat{\boldsymbol \Gamma}^{(2)}_{k+1}, \widehat{\boldsymbol D}_{k+1}) := \underset{\boldsymbol \Omega^{(2)}, \boldsymbol \Gamma^{(2)}, {\bf D}}{\operatorname{argmin}} \{\dfrac{\rho}{2} \| \boldsymbol \Omega^{(2)} - (\boldsymbol {\widehat\Omega}_k - {\bf {\widehat U}}_k^{(2)}/\rho)\|^2_F 
+ \dfrac{\rho}{2} \| \boldsymbol \Gamma^{(2)} - (\boldsymbol {\widehat\Gamma}_k - {\bf {\widehat U}}_k^{(5)}/\rho)\|^2_F \\ \text{s.t.}  \ 
\boldsymbol \Omega^{(2)} = {\bf A}\boldsymbol{\Gamma}^{(2)} + {\bf D}, \ 
{\bf D} \ \text{diagonal}, \ D_{jj} \geq 0 \ \text{for} \  j=1,\ldots,p ,
\} \label{ADMMOmega2} \nonumber 
\end{multline}

The solution 
\begin{equation}
\widehat{\boldsymbol \Omega}^{(2)}_{k+1} = {\bf A}\widehat{\boldsymbol \Gamma}^{(2)}_{k+1} + {\bf {\widehat D}}_{k+1} \label{Omega2} 
\end{equation} is immediate and we are left with 
\begin{equation}
(\widehat{\boldsymbol \Gamma}^{(2)}_{k+1}, \widehat{\boldsymbol D}_{k+1}) := \underset{\boldsymbol \Gamma^{(2)}, {\bf D}}{\operatorname{argmin}} 
\{\dfrac{1}{2} \| {\bf {\widetilde A}}\boldsymbol{\Gamma}^{(2)} + {\bf {\widetilde D}} - {\bf {\widetilde M}}\|^2_F 
\  \text{s.t.}  \ 
{\bf D} \ \text{diagonal}, \ D_{jj} \geq 0 \ \text{for} \  j=1,\ldots,p ,
\} \nonumber 
\end{equation}
where we have substituted  ${\boldsymbol \Omega}^{(2)} = {\bf A}{\boldsymbol \Gamma}^{(2)} + {\bf {D}}$
and we denote
$${\bf {\widetilde A}} = \begin{pmatrix}
{\bf A} \\ {{\bf I}_{|\mathcal{T}|}}
\end{pmatrix} \in \mathds{R}^{(p+ |\mathcal{T}|)\times |\mathcal{T}|}, \  {\bf {\widetilde D}} = \begin{pmatrix}
{\bf D} \\ {{\bf 0}_{|\mathcal{T}| \times p}}
\end{pmatrix} \in \mathds{R}^{(p+ |\mathcal{T}|)\times p}, \ \text{and} \ {\bf {\widetilde M}} = \begin{pmatrix}
\boldsymbol {\widehat\Omega}_k - {\bf {\widehat U}}_k^{(2)}/\rho \\ \boldsymbol {\widehat\Gamma}_k - {\bf {\widehat U}}_k^{(5)}/\rho
\end{pmatrix} \in \mathds{R}^{(p+ |\mathcal{T}|)\times p}.$$

The solution  
\begin{eqnarray} \boldsymbol {\widehat\Gamma}^{(2)}_{k+1}  & = &  ({\bf {\widetilde A}}^\top {\bf {\widetilde A}} )^{-1}{\bf {\widetilde A}}^\top ({\bf {\widetilde M}} - {\bf {\widetilde D}}_{k+1}) \nonumber \\
& = &  ({\bf A}^\top{\bf A} + {\bf I}_{|\mathcal{T}|})^{-1}({\bf A}^\top : {\bf I}_{|\mathcal{T}|})({\bf \widetilde M} - {\bf {\widetilde D}}_{k+1}) \label{Gamma2}
\end{eqnarray} is  immediate and we are left with
\begin{eqnarray}
\widehat{\boldsymbol D}_{k+1}  & := & \underset{{\bf D}}{\operatorname{argmin}} 
\{\dfrac{1}{2} \|({\bf {\widetilde M}} - {\bf {\widetilde D}})  - {\bf {\widetilde A}}({\bf {\widetilde A}}^\top {\bf {\widetilde A}} )^{-1}{\bf {\widetilde A}}^\top ({\bf {\widetilde M}} - {\bf {\widetilde D}})\|^2_F 
\  \text{s.t.}  \ 
{\bf D} \ \text{diag.}, \ D_{jj} \geq 0 \ \text{for} \  j=1,\ldots,p , 
\} \nonumber \\
 & = & \underset{{\bf D}}{\operatorname{argmin}} 
\{\dfrac{1}{2} \| ({\bf I}_{p+|\mathcal{T}|} - {\bf {\widetilde A}}({\bf {\widetilde A}}^\top {\bf {\widetilde A}} )^{-1}{\bf {\widetilde A}}^\top)({\bf {\widetilde M}} - {\bf {\widetilde D}})\|^2_F 
\  \text{s.t.}  \ 
{\bf D} \ \text{diag.}, \ D_{jj} \geq 0 \ \text{for} \  j=1,\ldots,p , 
\} \nonumber \\
  & = & \underset{{\bf D}}{\operatorname{argmin}} 
\{\dfrac{1}{2} \| {\bf B} - {\bf C}{\bf {D}}\|^2_F 
\  \text{s.t.}  \ 
{\bf D} \ \text{diag.}, \ D_{jj} \geq 0 \ \text{for} \  j=1,\ldots,p , 
\}
\nonumber 
\end{eqnarray}
with ${\bf B} = ({{\bf I}_{p+|\mathcal{T}|}} - {\bf {\widetilde A}}({\bf {\widetilde A}}^\top {\bf {\widetilde A}} )^{-1}{\bf {\widetilde A}}^\top){\bf {\widetilde M}} \in \mathds{R}^{(p+ |\mathcal{T}|)\times p}, \ {\bf C} = ({{\bf I}_{p}} : {{\bf 0}_{p \times |\mathcal{T}|}})^\top - {\bf {\widetilde A}}({\bf {\widetilde A}}^\top {\bf {\widetilde A}} )^{-1}{\bf {A}}^\top \in \mathds{R}^{(p+ |\mathcal{T}|)\times p}$. 
The solution is 
\begin{equation} 
\text{diag}({\bf {\widehat D}}_{k+1}) = \text{diag}({\bf C}^\top {\bf C})^{-1}\text{diag}({\bf B}^\top {\bf C})_+. \label{equationD}
\end{equation}

\subsection{Solving for $\boldsymbol \Omega^{(3)}$ \label{om3}}
Minimizing the augmented Lagrangian with respect to $\boldsymbol \Omega^{(3)}$ gives
\begin{eqnarray}
\widehat{\boldsymbol \Omega}^{(3)}_{k+1}  & := &   \underset{\boldsymbol \Omega^{(3)}}{\operatorname{argmin}} \{
\langle {\bf U}^{(3)}, {\boldsymbol \Omega}^{(3)} - {\boldsymbol \Omega} \rangle + \dfrac{\rho}{2} \|{\boldsymbol \Omega}^{(3)} - {\boldsymbol \Omega} \|^2_F  + \lambda_2 \|\boldsymbol \Omega^{-\text{diag}(3)}\|_1
\}  \nonumber\\
& = &   \underset{\boldsymbol \Omega^{(3)}}{\operatorname{argmin}} \{
\dfrac{\rho}{2} \| \boldsymbol \Omega^{(3)} - (\boldsymbol {\widehat\Omega}_k - {\bf {\widehat U}}_k^{(3)}/\rho)\|^2_F  + \dfrac{1}{2\rho} \|{\bf U}^{(3)}\|_F^2  + \lambda_2 \|\boldsymbol \Omega^{-\text{diag}(3)}\|_1
\}  \nonumber \\
& = &   \underset{\boldsymbol \Omega^{(3)}}{\operatorname{argmin}} \{
\dfrac{\rho}{2} \| \boldsymbol \Omega^{(3)} - (\boldsymbol {\widehat\Omega}_k - {\bf {\widehat U}}_k^{(3)}/\rho)\|^2_F + \lambda_2 \|\boldsymbol \Omega^{-\text{diag}(3)}\|_1
\} \nonumber  \label{ADMMOmega3}
\end{eqnarray}
The solution is simply elementwise soft-thresholding:
\begin{equation}
\widehat{\Omega}^{(3)}_{k+1, ij} =
\begin{cases}
S({\widehat\Omega}_{k, ij} - { {\widehat U}}_{k, ij}^{(3)}/\rho, \lambda_2/\rho),  & \text{if} \ i\neq j\\
{\widehat\Omega}_{k, ij} - { {\widehat U}}_{k, ij}^{(3)}/\rho , & \text{if} \  i=j,\\
\end{cases}
   \label{Omega3}
\end{equation}
with the soft-threshold operator $S(\omega, \lambda) = \text{sign}(\omega) \text{max}(|\omega| - \lambda, 0)$ applied to $\omega \in \mathds{R}$. 

\subsection{Update Variables $\boldsymbol \Omega$ and $\boldsymbol \Gamma$} \label{global}
Minimizing the augmented Lagrangian with respect to variables $\boldsymbol \Omega$ and $\boldsymbol \Gamma$
gives
\begin{eqnarray}
\widehat{\boldsymbol \Omega}_{k+1} & := & \underset{\boldsymbol \Omega}{\operatorname{argmin}} \left\{
\sum_{i=1}^3 \|\boldsymbol{\widehat{\Omega}}^{(i)}_{k+1} - (\boldsymbol{\Omega} - \widehat{{\bf U}}^{(i)}_{k}/\rho)\|_F^2 
\right\} = 
\bar{{\boldsymbol \Omega}}_{k+1} + \dfrac{1}{\rho} \bar{\bf U}^\Omega_{k} \label{globomega}  \\ 
\widehat{\boldsymbol \Gamma}_{k+1} & := & \underset{\boldsymbol \Gamma}{\operatorname{argmin}} \left\{
\sum_{i=1}^2 \|\boldsymbol{\widehat{\Gamma}}^{(i)}_{k+1} - (\boldsymbol{\Gamma} - \widehat{{\bf U}}^{(i+3)}_{k}/\rho)\|_F^2 
\right\} 
= \bar{{\boldsymbol \Gamma}}_{k+1} + \dfrac{1}{\rho} \bar{\bf U}^\Gamma_{k}, \label{globgamma} 
\end{eqnarray}
where $ \bar{{\boldsymbol \Omega}}_{k} := \dfrac{\widehat{\boldsymbol \Omega}^{(1)}_{k} + \widehat{\boldsymbol \Omega}^{(2)}_{k} + \widehat{\boldsymbol \Omega}^{(3)}_{k}}{3}, 
\bar{\bf U}^\Omega_{k} := \dfrac{\widehat{\bf U}^{(1)}_{k} + \widehat{\bf U}^{(2)}_{k} + \widehat{\bf U}^{(3)}_{k}}{3},
\bar{{\boldsymbol \Gamma}}_{k} := \dfrac{\widehat{\boldsymbol \Gamma}^{(1)}_{k} + \widehat{\boldsymbol \Gamma}^{(2)}_{k}}{2},  \bar{\bf U}^\Gamma_{k} := \dfrac{\widehat{\bf U}^{(4)}_{k} + \widehat{\bf U}^{(5)}_{k}}{2}.$

\subsection{Update Dual Variables} \label{dual}
The updates of the dual variables are given by
\begin{eqnarray}
\widehat{\bf U}^{(i)}_{k+1} & := &   \widehat{\bf U}^{(i)}_{k} + \rho \left ( \widehat{{\boldsymbol \Omega}}_{k+1}^{(i)} - \widehat{{\boldsymbol \Omega}}_{k+1} \right), \ \text{for} \ i=1, \ldots, 3\nonumber \\
\widehat{\bf U}^{(j+3)}_{k+1} & := & \widehat{\bf U}^{(j+3)}_{k} + \rho \left ( \widehat{{\boldsymbol \Gamma}}_{k+1}^{(j)} - \widehat{{\boldsymbol \Gamma}}_{k+1} \right), \ \text{for} \ j=1, \ldots, 2.  \nonumber 
\end{eqnarray}
Similarly, averaging the first three updates  and the latter two gives
\begin{eqnarray}
\bar{\bf U}^\Omega_{k+1} & := &   \bar{\bf U}^\Omega_{k} + \rho \left ( \bar{{\boldsymbol \Omega}}_{k+1} - \widehat{{\boldsymbol \Omega}}_{k+1} \right), \ \text{for} \ i=1, \ldots, 3 \label{dualomega} \\
\bar{\bf U}^\Gamma_{k+1} & := & \bar{\bf U}^\Gamma_{k} + \rho \left ( \bar{{\boldsymbol \Gamma}}_{k+1} - \widehat{{\boldsymbol \Gamma}}_{k+1} \right), \ \text{for} \ j=1, \ldots, 2, \label{dualgamma}
\end{eqnarray}
Substituting \eqref{globomega} and \eqref{globgamma} into \eqref{dualomega} and \eqref{dualgamma} yields that $\bar{\bf U}^\Omega_{k+1} = \bar{\bf U}^\Gamma_{k+1} = {\bf 0}$ after the first iteration.

\newpage

\section{Additional Simulation Results} \label{Appendix_sims}

\begin{figure}[H]
	\centering
\includegraphics[width=0.45\textwidth]{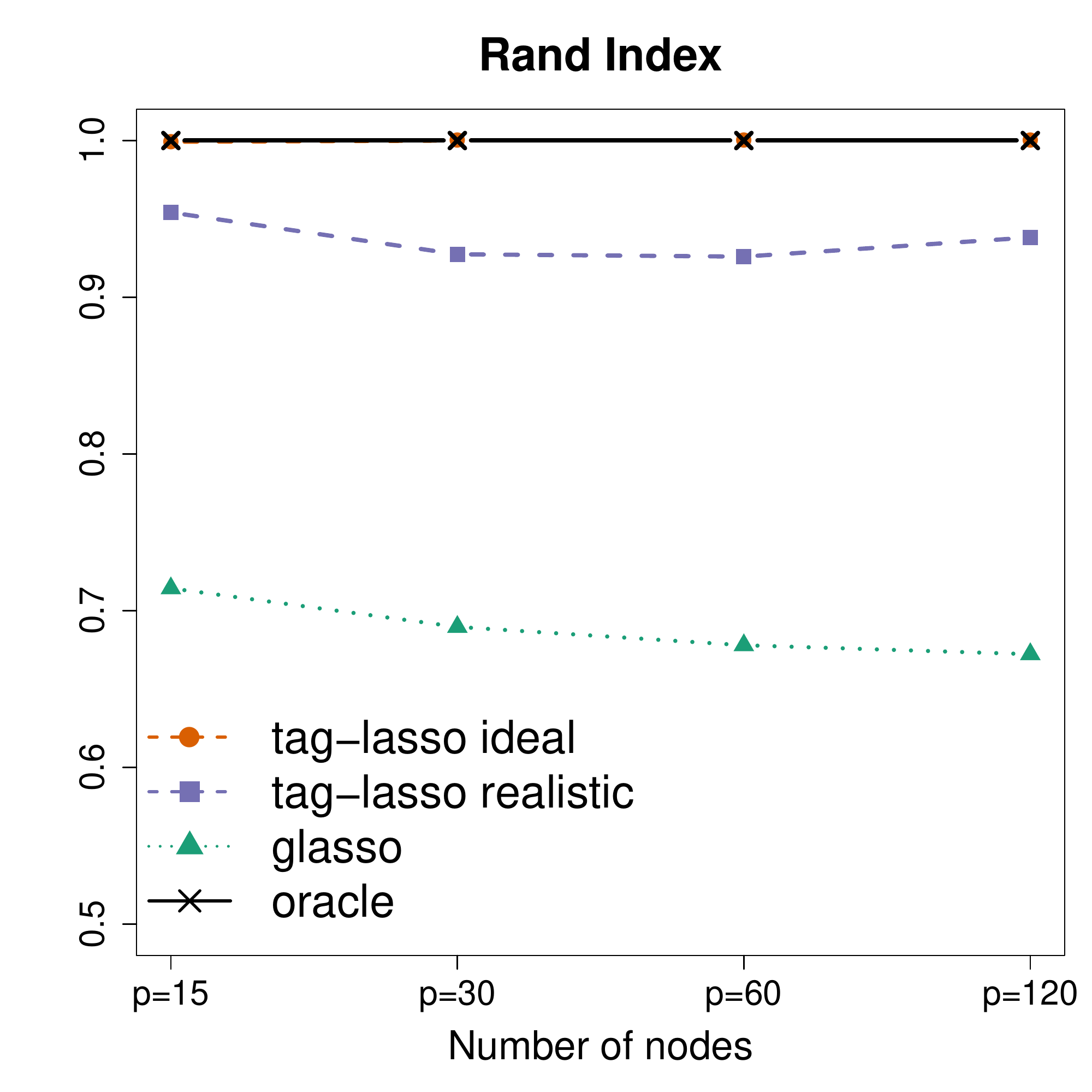}
\hspace{1cm}
\includegraphics[width=0.45\textwidth]{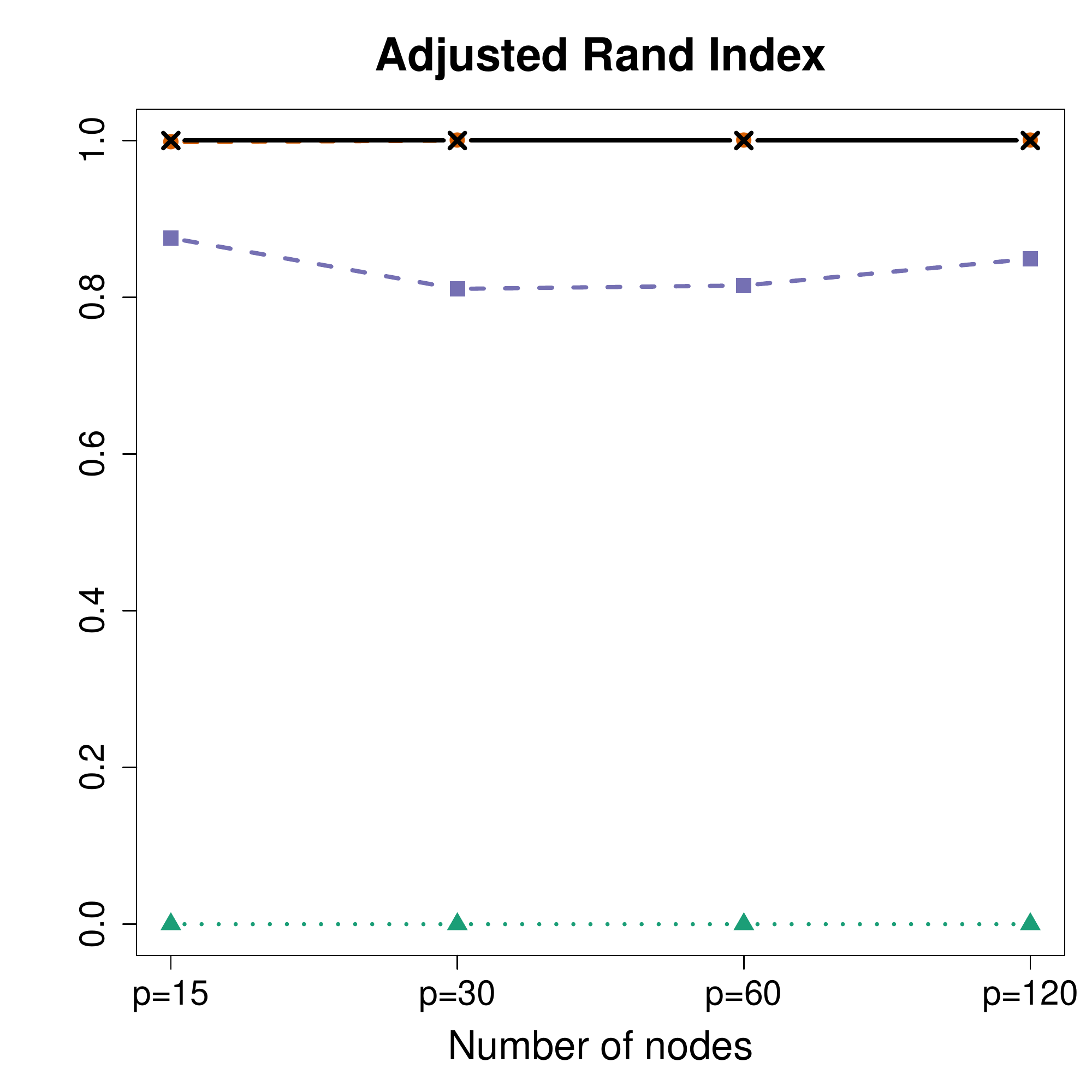}
\includegraphics[width=0.45\textwidth]{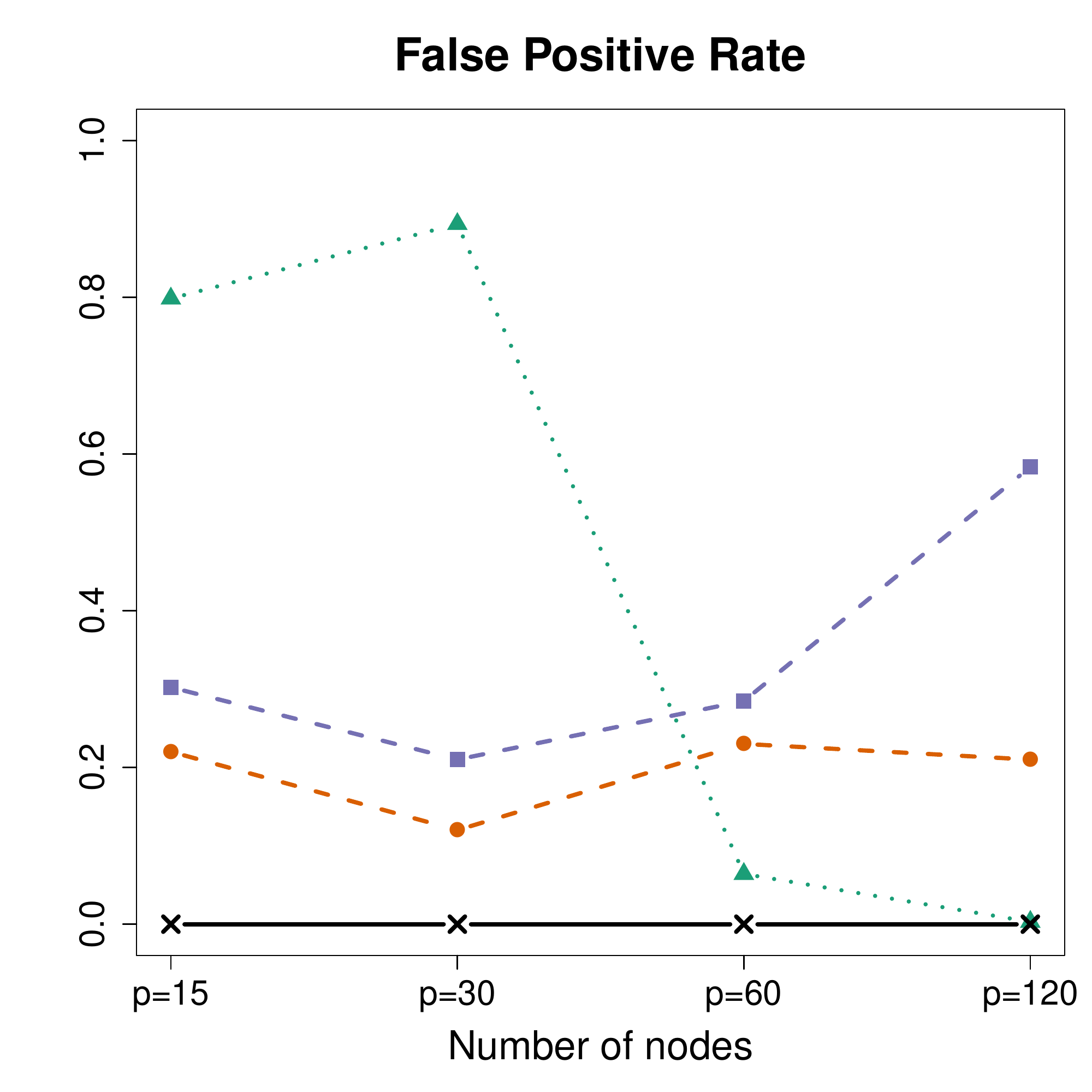}
\hspace{1cm}
\includegraphics[width=0.45\textwidth]{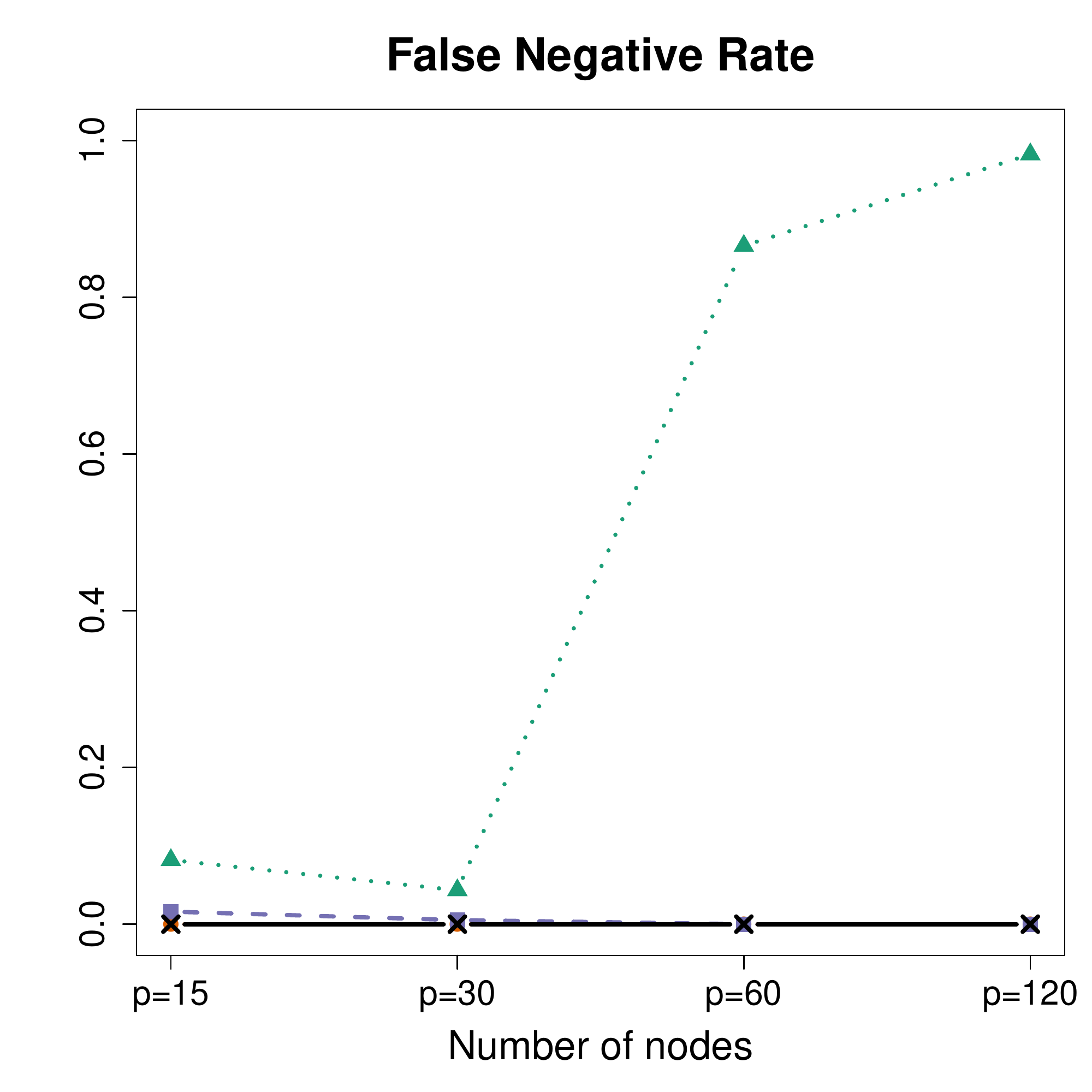}
\caption{ \label{sims_increasep_agg_and_sparsity} Simulation results for increasing number of nodes $p$.
Top: Aggregation performance (RI: left; ARI: right);
Bottom: Sparsity recovery (FPR: left; FNR: right) of the four estimators}
\end{figure}

\begin{figure}
	\centering
\includegraphics[width=0.45\textwidth]{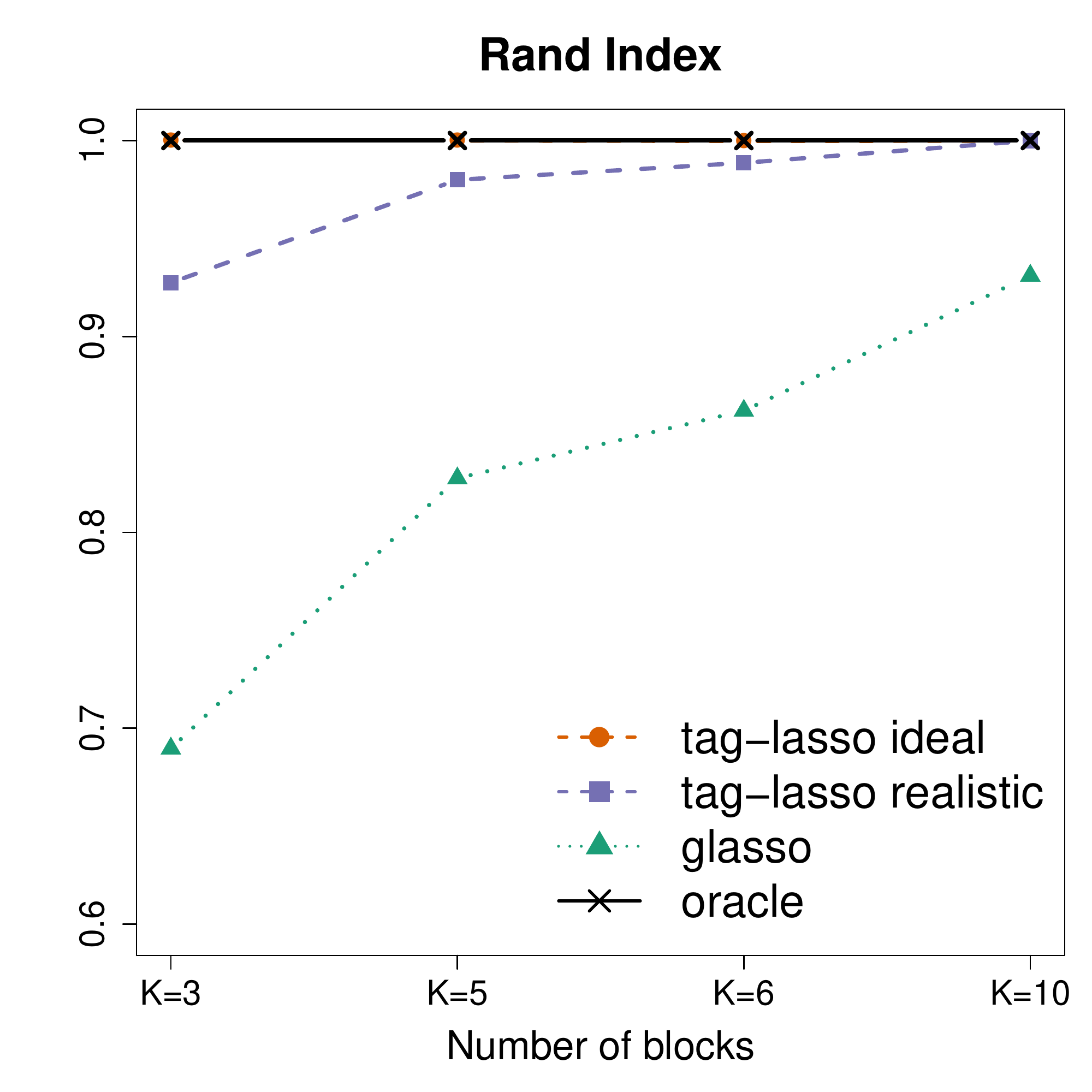}
\hspace{1cm}
\includegraphics[width=0.45\textwidth]{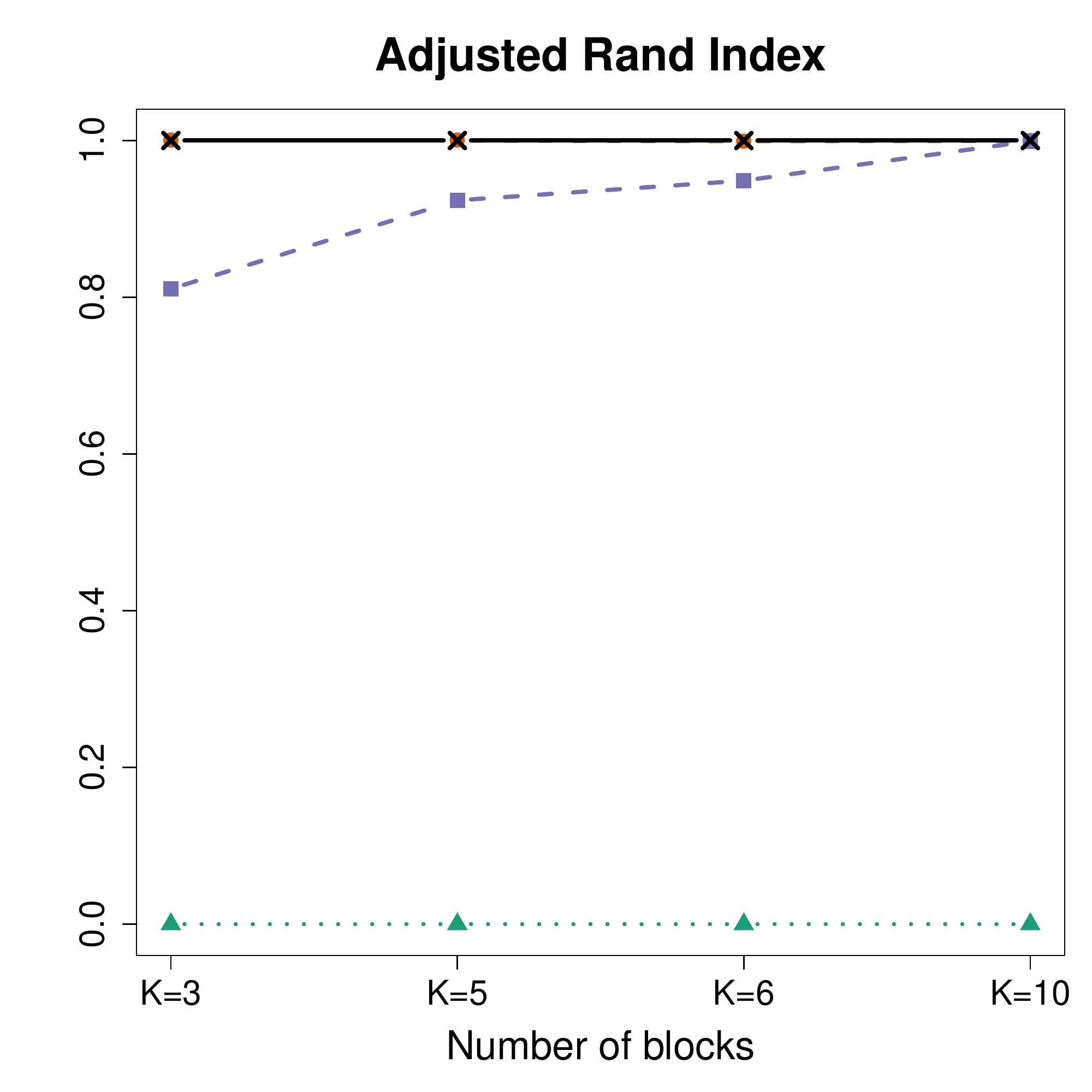}
\includegraphics[width=0.45\textwidth]{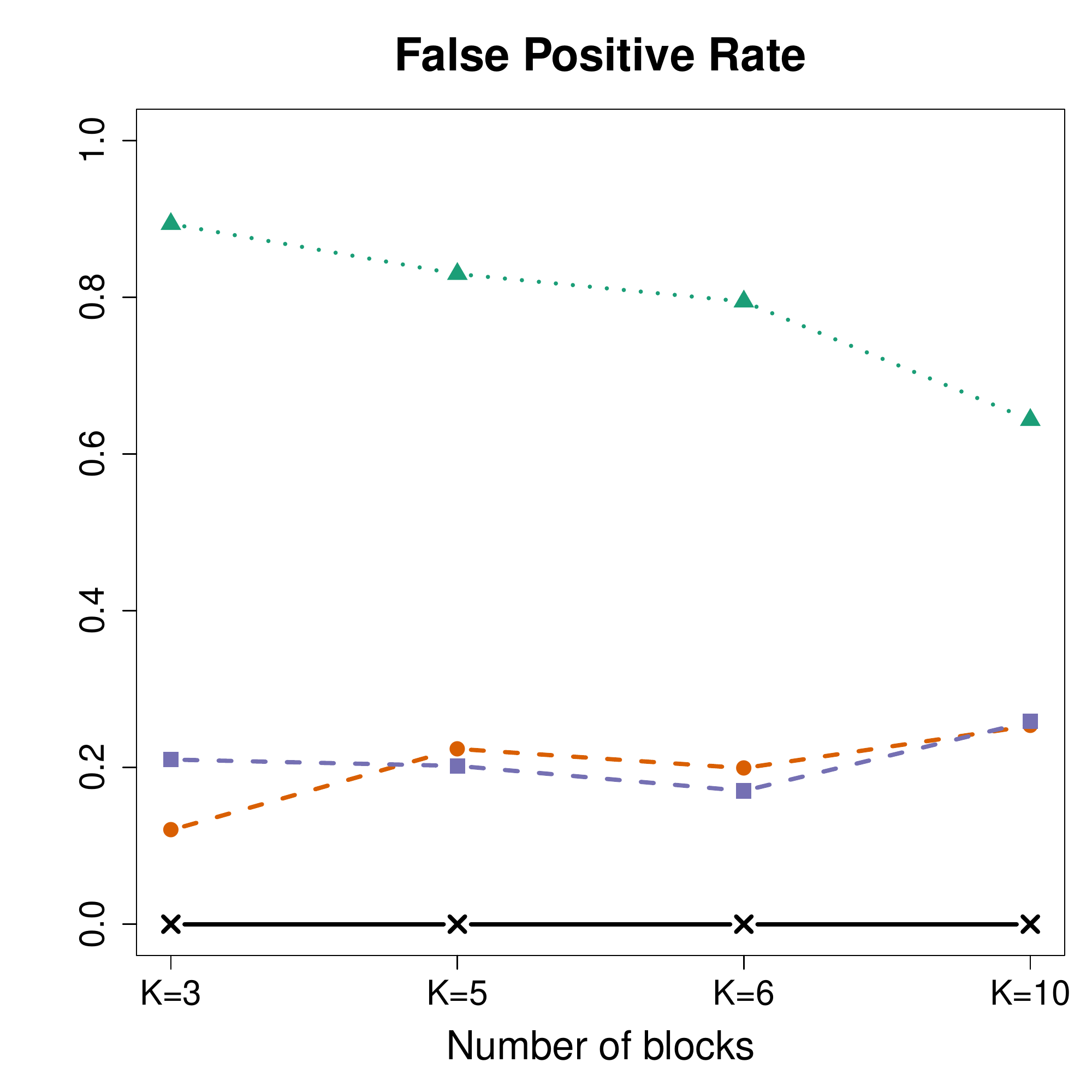}
\hspace{1cm}
\includegraphics[width=0.45\textwidth]{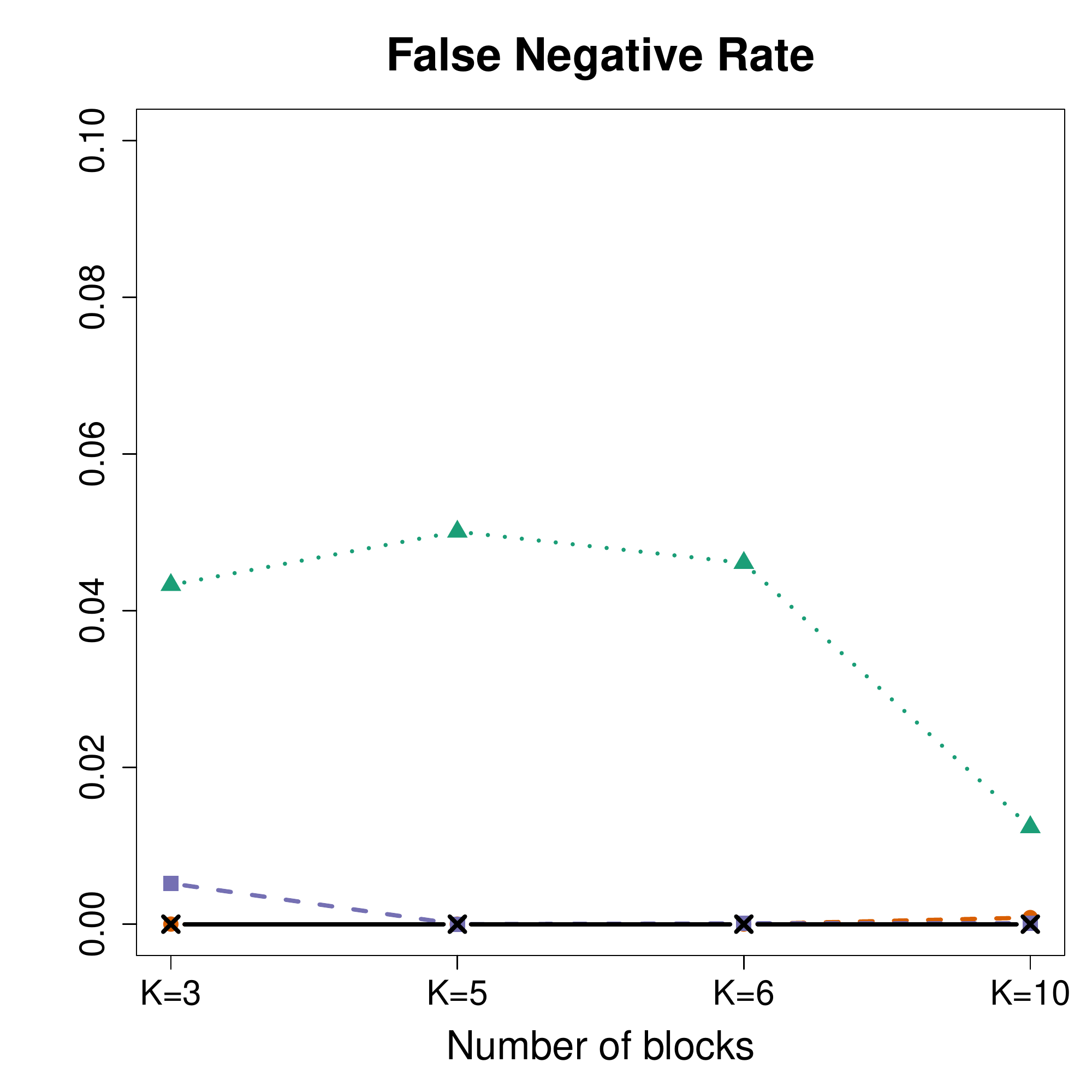}
\caption{ \label{sims_increasek_agg_and_sparsity} Simulation results for increasing number of blocks $K$.
Top: Aggregation performance (RI: left; ARI: right);
Bottom: Sparsity recovery (FPR: left; FNR: right) of the four estimators}
\end{figure}

\newpage
\section{Financial Application: Data Description} \label{Appendix_Finance}
	
\begin{table}[H]
\centering
\caption{Financial Application: Data Description, as taken from https://realized.oxford-man.ox.ac.uk/data/assets.}
\begin{tabular}{lll}\hline 
Abbreviation & Description & Location \\ \hline 
DJI	&Dow Jones Industrial Average&	US \\
IXIC&	Nasdaq 100&	US \\
SPX &	S\&P 500 Index	&US \\
RUT	&Russel 2000	&US \\
GSPTSE& 	S\&P/TSX Composite index&	Canada \\

BVSP&	BVSP BOVESPA Index	&Brazil \\
MXX	&IPC Mexico	&Mexico \\

OMXC20	&OMX Copenhagen 20 Index&	Denmark \\
OMXHPI	&OMX Helsinki All Share Index&	Finland \\
OMXSPI	&OMX Stockholm All Share Index&	Sweden \\
OSEAX	&Oslo Exchange All-share Index&	Norway \\

GDAXI &	Deutscher Aktienindex	& Germany \\
SSMI	&Swiss Stock Market Index	&Switzerland \\

BVLG&	Portuguese Stock Index & Portugal \\
FTMIB &	Financial Times Stock Exchange Milano Indice di Borsa	&Italy \\
IBEX&	Iberia Index 35 &	Spain \\
SMSI	& General Madrid Index&	Spain \\

AEX	&Amsterdam Exchange Index	&Netherlands \\
BFX	&Bell 20 Index&	Belgium \\
FCHI &	Cotation Assist\'ee en Continue 40	&France \\
FTSE	&Financial Times Stock Exchange 100	&UK \\

STOXX50E&	EURO STOXX 50	&Europe \\

HSI&	HANG SENG Index	& Hong Kong \\
KS11&	Korea Composite Stock Price Index (KOSPI)&	South Korea \\
N225&	Nikkei 225&	Japan \\
SSEC&	Shanghai Composite Index&	China \\
STI	&Straits Times Index&	Singapore \\

KSE	&Karachi SE 100 Index	&Pakistan \\
BSESN &	S\&P Bombay Stock Exchange Sensitive Index	 & India \\
NSEI	&NIFTY 50&	India \\

AORD&	All Ordinaries Index&	Australia \\

\hline

\end{tabular}
\end{table}
\end{document}